\documentclass[aps,prb,reprint,twocolumn, superscriptaddress,longbibliography]{revtex4-1}
\usepackage{bbold}
\usepackage{amsmath,amsfonts,amsmath,mathtools,physics,bbm,bm}
\usepackage{graphicx}
\usepackage{marginnote}
\usepackage{braket}
\usepackage[colorlinks,linkcolor=blue,urlcolor=blue,citecolor=blue]{hyperref}
\usepackage[all]{hypcap}

\usepackage{blindtext}
\usepackage{enumitem}
\usepackage{xcolor}
\usepackage[normalem]{ulem}
\usepackage{float}
\usepackage{dsfont}
\setcitestyle{numbers,square}

\newcommand{\SFF}{{\mathrm{SFF}}}
\newcommand{\tH}{t_{\mathrm{H}}}

\definecolor{darkblue}{rgb}{0.0, 0.0, 0.60}

\def\Re{\text{Re}}
\def\({\left (}
\def\){\right )}

\graphicspath{{Figure/PNG/}{Figure/PDF/}{Figure/EPS/}{Figure/TEX/}{Figure/}}

\newcommand\mydots{\hbox to 1em{.\hss.\hss.}}

\begin{document}

\title{
Statistics of the Random Matrix Spectral Form Factor
}

\author{Alex Altland}
\affiliation{Institut f\"ur Theoretische Physik, Universit\"at zu K\"oln, Z\"ulpicher Straße 77, 50937 K\"oln, Germany}

\author{Francisco Divi}
\affiliation{Perimeter Institute for Theoretical Physics, Waterloo, Ontario N2L 2Y5, Canada}
\affiliation{ICTP South American Institute for Fundamental Research, Instituto de F\'{i}sica Te\'{o}rica,\\
Univ. Estadual Paulista, Rua Dr. Bento Teobaldo Ferraz 271, 01140-070, S\~{a}o Paulo, Brazil} 

\author{Tobias Micklitz}
\affiliation{Centro Brasileiro de Pesquisas Físicas, Rua Xavier Sigaud 150, 22290-180, Rio de Janeiro, Brazil}

\author{Silvia Pappalardi}
\affiliation{Institut f\"ur Theoretische Physik, Universit\"at zu K\"oln, Z\"ulpicher Straße 77, 50937 K\"oln, Germany}

\author{Maedeh Rezaei}
\affiliation{Institut f\"ur Theoretische Physik, Universit\"at zu K\"oln, Z\"ulpicher Straße 77, 50937 K\"oln, Germany}

\date{\today}

\begin{abstract}
The spectral form factor of random matrix theory plays a key role in the
description of disordered and chaotic quantum systems. While its moments are
known to be approximately Gaussian,  corrections subleading in the matrix dimension, $D$, have recently come to 
attention, with conflicting results in the literature. In this work, we
investigate these departures from Gaussianity for both circular and Gaussian
ensembles. Using two independent approaches ---  
sine-kernel techniques and supersymmetric field theory --- we identify the form
factor statistics to next leading order in a $D^{-1}$ expansion. Our
sine-kernel analysis highlights inconsistencies with previous studies, while the
supersymmetric approach backs these findings and suggests an understanding of
the statistics from a complementary perspective. Our findings fully agree with
numerics. They are presented in a pedagogical way, highlighting pathways (and pitfalls) in the study of statistical signatures at next
leading order, which are increasingly becoming important in applications.  
\end{abstract}

\maketitle

\section{Introduction}

The spectral form factor 
\begin{align}
    \label{eq:DefinitionSFF}
    \SFF\equiv |\tr U_t|^2
\end{align}
is one of the most important observables in the
characterization of disordered or chaotic quantum systems \cite{berry1985semiclassical,sieber2001correlations, mueller2004semiclassical, mueller2005periodic}. It owes its
significance to a combination of factors: The SFF probes
the dynamics of the time evolution operator $U_t$, and its Fourier transform
energy dependent correlations in spectra, both key signatures of a system. The time dependence of
numerous observables, such as correlation functions between operators
\cite{delacretaz2020heavy, yoshimura2023operator, bouverotdupuis2025randomm},
out of time order correlations (OTOC) functions \cite{cotler2020spectral,
pappalardi2024eigenstate}, or return probabilities \cite{torres2018generic},
approximately, or even exactly reduce to the  SFF. Finally, the SFF has come to
prominence in the context of many-body physics
\cite{chan2018solution, chan2018spectral, kos2018many, bertini2018exact, altland2018quantum} and
holography \cite{garcia2016spectral, cotler2017black, cotler2017chaos, 
 saad2019late, saad2019semiclassical,
altland2021from}, where it features in the effective boundary description of
two-dimensional gravitational theories.

The spectral form factor is commonly discussed with reference to three temporal
regimes: at times $t<t_\textrm{Th}$ smaller than an effective \emph{Thouless
time}, it probes physical relaxation processes preceding the entry into an
ergodic regime at times $t>t_\textrm{Th}$. (Our focus here will be on systems
with late time ergodic behavior.) Once ergodic, it
typically increases in a linear or near linear manner --- the so called `ramp
regime' --- up to a time ${t_\textrm{H}=2\pi/\Delta}$, the \emph{Heisenberg time}
defined as the inverse of a system's characteristic level spacing $\Delta$. At later
times, the form factor approaches constant behavior, the `plateau regime'. Our
focus in this paper will be on ergodic late time physics, ramp and plateau.

Even in the ergodic regime, the spectral form factor remains a statistically distributed quantity,
and is subject to three
different sources of fluctuations:
\begin{enumerate}
    \item The SFF of generic chaotic quantum systems typically shows deviations
    from that of random matrix theory (RMT), even deep in the ergodic regime at times ${t\simeq
    t_\textrm{H}}$. These deviations are due to large sample-to-sample
    fluctuations of the spectral density, which in many-body systems generally
    exceed those of  Gaussian matrix ensembles. The deviations from the RMT form
    factor caused by these fluctuations have become a subject of study in their
    own right \cite{flack2020statistics, chan2021spectral, winer2022spectral,
    legramandi2024moments, fritzsch2024universal}. 
    \item The form factor makes reference to physical time so that, by
    energy/time uncertainty, it includes an integral over energies (cf.
    Eqs.~\eqref{eq_main} and \eqref{eq_zt} below). The non-constancy of the
    spectral density of a quantum system affects the form factor and its
    statistical moments via this convolution mechanism \cite{brezin1997spectral}. 
    \item Even within RMT, the spectral form factor is a famously non self-averaging quantity
    whose statistical fluctuations are of about the same order as the average,
    and not diminishing in the large system size limit \cite{prange1997spectral, braun2015self, matsoukas2023unitarity}.   
\end{enumerate} 
In generic interacting quantum systems, these three fluctuation channels act in
combination, leading to intricate statistics. Their importance is highlighted by
the fact that the quantum systems relevant to present day applications are
typically large, $D \gg 1$, but not thermodynamically large. Some of the
fluctuation channels mentioned above diminish in the large system size limit,
$D\to \infty$, but not all of them do. Finally, statistical data on the form
factor is typically collected by sampling over a finite number of runs
$N_\textrm{sim}$, either on a computer or in experiments. This opens a fourth
fluctuation channel reflecting the statistics of a finite data set. 
In view of this level of complexity, it is important to understand the statistics of the
random matrix SFF as a `null-model' before comparing it to realistic quantum
systems.

The statistics of the form factor in RMT has been the subject of a number of
previous studies \cite{haakeSommers1999, cotler2017chaos, Liu2019spectral,
forrester2021quantifying, forrester2021differential, cipolloni2023spectral},  a
key finding being that it is approximately Gaussian, which at the second order
implies $\overline{\mathrm{SFF}^2} \simeq 2(\overline{\mathrm{SFF}})^2$, where
$\overline{\,\cdot\, }$ is the ensemble average. Deviations from Gaussianity are
subleading in $D$ but otherwise universal signatures of the quantum ergodic
phase. They, too, have come under scrutiny, with a plethora of not
entirely consistent
results: Ref.~\cite{cipolloni2023spectral}  rigorously computes
$\overline{\mathrm{SFF}}$ and $\text{Var (SFF)}$ for Wigner matrices, but only
for short times $t\ll N^{5/17}$, where it is dominated by the edge spectrum. 
Ref.~\cite{haakeSommers1999} agrees with
Ref.~\cite{cipolloni2023spectral} when such edge effect is neglected, but it
also predicts the emergence of non-Gaussian statistics linearly increasing in
time, $t$, at $t=\tH/2$. The sudden (non-analytic) onset of this correction is
indicative of a non-perturbative effect, similar to the celebrated
`ramp-plateau' transition in the average SFF. Finally, Refs.
\cite{cotler2017chaos, Liu2019spectral} predict a more complicated time
dependence containing a linear term throughout the semiclassical regime
$t<\tH/2$, and contributions of next-leading order in the $1/D$-expansion. (We
note that for \cite{cotler2017chaos}, this equation is ancillary to the paper's
main results.)

In the comparison of these results, it is important to remember that the SFF is
subject to energy averaging, the second channel of fluctuations mentioned above:
The SFF contains physical time in the dimensionless combination $\tau \equiv
t\Delta/2\pi$, where $\Delta\sim D^{-1}$ is the characteristic level spacing. Being
defined in terms of an unconstrained trace $\SFF=|\textrm{tr}(U_t)|^2$ the scale
$\Delta =\Delta(\epsilon)$ enters averaged over the entire spectrum blurring otherwise
sharply defined structures over scales $\sim\Delta^{-1}$ \cite{brezin1997spectral}. This mechanism is
absent in \emph{circular ensembles}, defined in terms of Haar-distributed unitary
operators with statistically uniform eigenphase spacing, $\Delta=2\pi/D$, making
them attractive candidates for the study of the more interesting
`intrinsic' statistical fluctuations, channel 3.

\begin{figure}[t]
	\centering
\includegraphics[width=1 \linewidth]{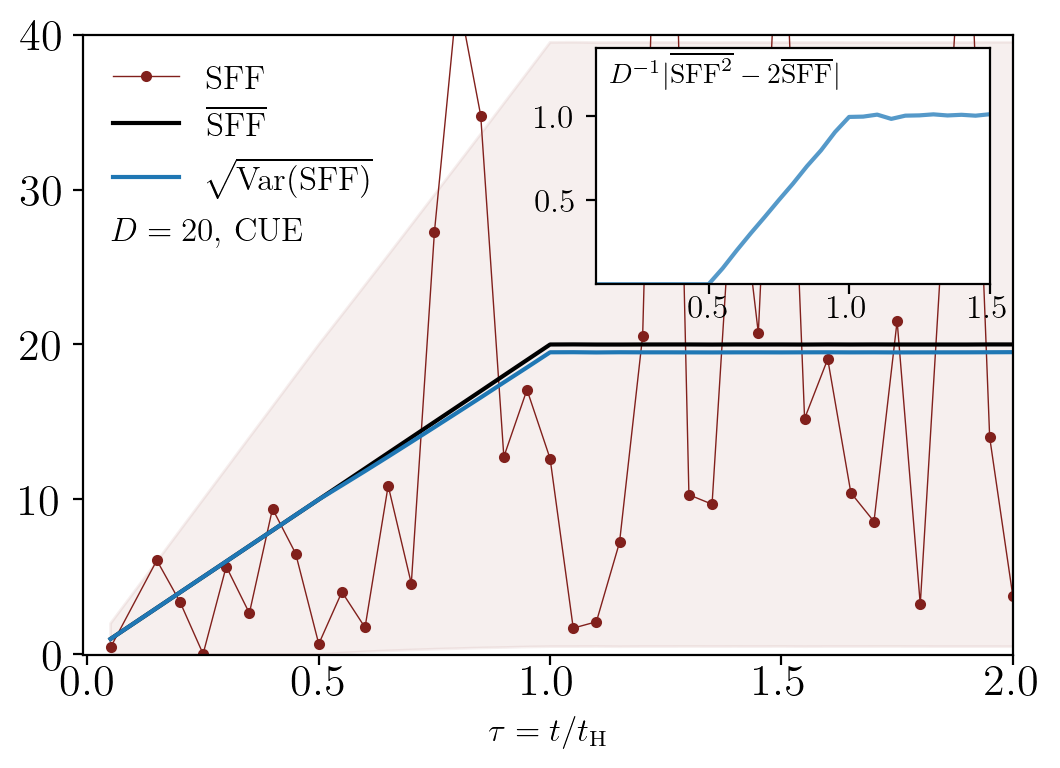}
	\caption{Numerical summary of the paper. The spectral form factor (SFF) of a single realization (red dots) fluctuates over time around its average $\overline{\rm SFF}$ (black line), with a variance ${\rm Var(SFF)}=\overline{\rm SFF^2}-\overline{\rm SFF}^2$ proportional to the average in the large-$D$ limit (blue line), i.e. Gaussian distribution. The amplitude of these fluctuations is indicated by the shaded red region. The data show deviations from the Gaussian behavior for $\tau = t/t_{\rm H} > 1/2$, as emphasized in the inset, which shows $D^{-1}|\overline{\SFF^2}-2\overline{\SFF}^2|$. This work addresses precisely such deviations from Gaussianity, providing two analytical frameworks to account for them. Numerical simulation from a CUE for $D=20$, average with $N_{\rm sim}=2 \cdot 10^7$ samples.}
	\label{fig_0}
 \vspace{-.5cm}
\end{figure}

In this paper, we analyze the statistics of the SFF in random matrix theory and
discuss the departures from Gaussianity, which are non-perturbative and
universal. We apply a combination of different methods --- random matrix
techniques, field theory, diagrammatic perturbation theory, and direct
diagonalization --- to analyze the statistics, emphasis put on the variance. Our
numerical analysis focuses on models with relatively small $D$, where computing
power is channeled into runs over many realizations and the deviations from
Gaussian statistics are clearly resolved, as shown in Fig.~\ref{fig_0}, where we
exemplify our results.  \\
As summarized below, we analytically confirm and numerically validate the
results of Ref. \cite{haakeSommers1999} for circular ensembles. Indeed,
the results of that reference follow by Fourier transformation from mathematically
rigorous results for CUE spectral
determinants~\cite{conrey2007howepairssupersymmetryratios}, and on this basis are
exact. We demonstrate that a conflicting formula for $\overline{\SFF^2}$ in  Refs.~\cite{cotler2017chaos,
Liu2019spectral} results from computational mistakes. Referring to the
mathematical Ref. \cite{cipolloni2023spectral}, which critically reviewed  common analytical approaches to random hermitian matrices, we qualitatively discuss our
understanding of their status in the analysis of $1/D$-subleading signatures
to the SFF statistics.\\

In the next section, we provide a brief summary of our results. The rest of the paper is organized by technique: in Sec.~\ref{sec_RMT}, we use the sine-kernel approach of random matrix theory. After a review of the method and notations, we compute the second moment of the spectral form factor, emphasizing the differences with the formula for the second moment in Refs.~\cite{cotler2017chaos, Liu2019spectral}. In Sec.~\ref{sec_fieldtheory}, we use the supersymmetric method, employing a two-fold replicated non-linear sigma model.

\subsection{Summary of results}

In this paper, we study the statistical moments of the spectral form factor
${\SFF} = |\tr U_t |^2$ where $U_t$ is a $D\times D$ matrix sampled from a
random matrix theory ensemble. It is well known that in the large $D$ limit, the
connected part of the spectral form factor, i.e.
\begin{align*}
    {  \SFF_c} = |\tr U_t|^2-|\overline{\tr U_t}|^2
\end{align*}
 contains universal properties about the spectral correlations of the level
density $ \rho(\lambda) = \sum_{i=1}^D\delta(\lambda-\lambda_i)$ where
$\{\lambda_i\}_{i=1, \dots D}$ are the eigenphases of the unitary generating the
dynamics $U_t$. For definiteness, we focus on the unitary case and consider: I)
evolution generated by the Circular Unitary Ensemble (CUE), i.e., the
case $U_t = U^t$, where $t\in \mathbb N$ is discrete-time and  $U$ is
sampled uniformly from the unitary group;  II) $U_t=e^{-i H t}$, where $H$ is
sampled according to the Gaussian Unitary Ensemble (GUE).

\medskip 
The statistical properties of the SFF in the ergodic regime can be summarized via its moments as follows

\begin{equation}
    \label{eq_main}
    \overline{{ (\SFF_c)}^n} = D^n \left (n!  z(t) ^n + \frac 1 {D} \text{Conn}_n(t) + \mathcal O\left ( \frac{{z(t)^n}
    }{\sqrt{N_{\rm sim}}}\right ) \right).
\end{equation}
Here, the leading term is given by the Gaussian contribution, expressed
 in terms of the average of the SFF, denoted $z(t)$, which is given by the
 Fourier integral
\begin{equation}
\label{eq_zt}
    z(t) = \frac 1D \int d\lambda_1 d\lambda_2 \, \bar \rho_c(\lambda_1, \lambda_2)~e^{i(\lambda_1-\lambda_2)t} \ ,
\end{equation}
where $\bar \rho_c(\lambda_1, \lambda_2)=\overline{\rho(\lambda_1)
\rho(\lambda_2)}- \overline \rho(\lambda_1) \overline \rho(\lambda_2)$ is the
connected part of the two-point correlation function of the spectral density.
Specifically, the SFF of the  CUE with its constant density of states shows the
celebrated ramp-plateau behavior \cite{haakeSommers1999}
\begin{equation}
\label{eq_haake}
z(t) = \begin{dcases}
    t/t_{\rm H} \quad&\text{for } t<t_{\rm H}\\
    1 \qquad& \text{for } t\geq t_{\rm H}  \end{dcases}    \ .
\end{equation}
For the GUE, this result holds only approximately. Taking the non-constant
spectral density into account leads to the modified result, also valid for finite $D$, that for $t<\frac 4 \pi t_{\rm H}$ reads \cite{brezin1997spectral}
\begin{equation}
\label{eq_brezin}
z(t) = \frac 12 \frac t{t_{\rm H}}\sqrt{1-\left (\frac {\pi}{4}\frac t{t_{\rm H}}\right )^2} + 
\frac{2}{\pi}\arcsin{ \left ( \frac {\pi}{4}\frac t{t_{\rm H}}\right )} \ ,
\end{equation}
while it is one otherwise. \\

Secondly, the term $\mathcal O(...)$ indicates the error resulting from an
average over a finite number of simulations  $N_{\rm sim}$. It is
of the order of $z(t)^n$, since $\SFF$ is approximately Gaussian (the variance of
$\SFF^n$ scales as the square of its average), and it is suppressed as $N_{\rm
sim}^{-1/2}$ due to the central limit theorem.\\

Finally,  ${\rm Conn}_n(t)$, denotes the subleading connected
contribution to the higher moments of the spectral form factor (it vanishes for
$n=1$). Its computation for $n=2$ is the main result of the present study. \\

\begin{figure}[t]
	\centering
\includegraphics[width=1 \linewidth]{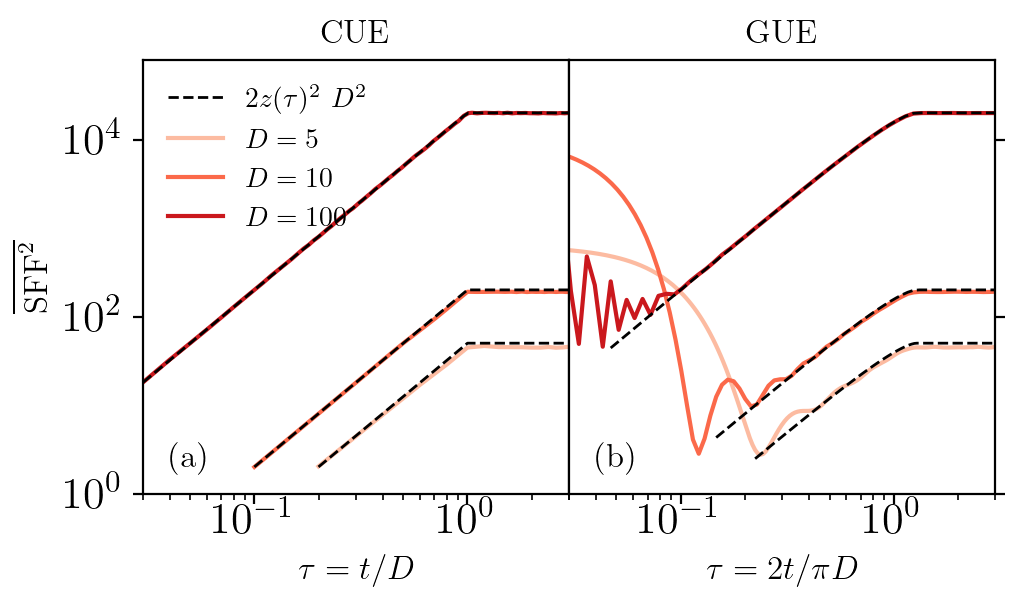}
\includegraphics[width=1 \linewidth]{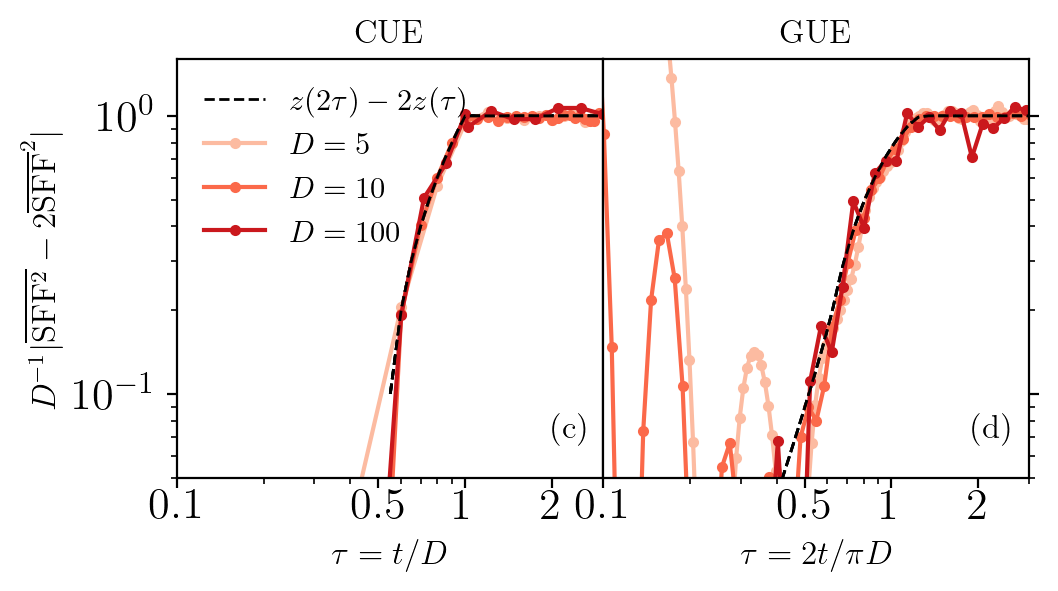}
	\caption{Leading (a-b) and subleading (c-d) behavior of the second moment of the spectral form factor. Results were sampled from the CUE (left) and for the evolution of a GUE Hamiltonian (right). The numerical data for $D=5, 10, 100$ (red lines with increasing color intensity) are compared with the analytical prediction [$2 z(\tau)^2$ (a-b) and $2z(\tau)-z(2\tau)$ (c-d)] (dashed black) using Eq. \eqref{eq_haake} and Eq. \eqref{eq_brezin} in the two panels. The empirical average is obtained with $N_{\rm sim}=(5 \cdot 10^4, 5 \cdot 10^5, 2 \cdot 10^6)$ with $D=5, 10, 100$ respectively. }
	\label{fig_1}
 \vspace{-.5cm}
\end{figure}

The rest of the paper is summarized as follows:
\begin{enumerate}
    \item We review how, to leading order in $D$, the SFF shows approximately
    Gaussian statistics, i.e. $ \overline{\SFF ^n} = n!  z(t) ^n$. This also
    determines the size of the oscillatory error $\mathcal O(...)$, due to the
    sample-to-sample fluctuations in numerical averages. We show how this
    asymptotic Gaussian result can be understood intuitively, and in terms of
    straightforward perturbation theory.
    \item We compute the subleading corrections at finite $D$ by focusing on $n=2$. They read
    \begin{equation}
        \label{eq_conn}
        {\rm Conn}_2(t) = z(2t) - 2 z(t)\ .
    \end{equation}
    When considering the CUE, this result reduces to a linear regime between
     $t_{\rm H}/2$ and $t_{\rm H}$ \cite{haakeSommers1999}, thus reproducing the
     data in the inset of Fig.\ref{fig_0}. Eq. \eqref{eq_conn} will follow from
     the application of two complementary  analytical approaches:
    \begin{itemize}
        \item Random matrix theory provides an 
        explicit result for  higher-order spectral correlations in terms of a
        determinant of the so-called sine-kernel, which allows an explicit
        computation of the second moment of the spectral form factor, and in
        particular the subleading connected correction in
        Eq. \eqref{eq_conn}. This approach has  been previously pursued in
        Refs.\cite{cotler2017chaos, Liu2019spectral} with, however, some
        computational errors affecting the formula for $\overline{\SFF^2}$.
        \item We apply the \emph{supersymmetric sigma model}, four-fold
        replicated to compute the fourth moment of the density of states. The
        integration manifold featuring in this framework contains a discrete
        number of saddle points, which are not accessible in perturbation theory
      and are responsible for the non-Gaussian contributions. 
        \end{itemize}
    \item These findings are confirmed numerically, as shown in Fig.~\ref{fig_1}
    for the leading and subleading behavior, in the two ensembles CUE and GUE.
    The plots show numerical averages obtained for large values of $N_{\rm
    sim}$, such that the sampling fluctuations are already suppressed, as
    discussed in Sec.~\ref{subsub_variance}. 
\end{enumerate}

\section{Random Matrix Theory}
\label{sec_RMT}
In this section, we review the statistics of the SFF, cf. Eq.~\eqref{eq_main},
as described by techniques of Random Matrix Theory. Specifically, we will
consider evolution generated by two standard ensembles: \\
\indent I) The Circular Unitary Ensemble (CUE), where
\begin{equation}
    U_t = U^t \quad \text{with} 
    \quad U\sim {\rm uniform} \quad t\in \mathbb N\ ,
\end{equation} 
$U$ being sampled uniformly from the unitary group. In this case, the
    eigenphases of $U$, $\lambda_i$,  are uniformly distributed over the unit
    circle, leading to a constant density of states $\overline{\rho}(\lambda) =
    \frac{D}{2\pi}$, such that $t_{\rm H}= 2 \pi  \bar \rho = D.$\\ \\
\indent II) The Gaussian Unitary Ensemble (GUE), where
\begin{align}
    \label{eq:GaussianStatistics}
    U_t = e^{- i H t}\quad  {\rm with}\quad P(H) \propto \exp({ -\frac{D}{2} \Tr(H^2)})\ .
\end{align}
The average density of states here is $\bar \rho(\lambda)
= \frac{D}{2\pi}\sqrt{4-\lambda^2}$, so one has $D$ eigenvalues in an interval
of size $4$, and the mean level density can be estimated as $\bar \rho = D/4$
and $t_{\rm H}=D \frac \pi 2$ \footnote{Note that this is not the only possible
estimate for $t_{\rm H}$. In fact one can consider $\bar \rho =
\overline\rho({\lambda=0})=D/\pi$ leading to $t_{\rm H}= 2D$, see e.g.
Refs.\cite{cotler2017chaos, Liu2019spectral} However, we find that $\bar \rho=D/4$ ($t_{\rm H}= D
\frac \pi2$) is more accurate in reproducing the numerical results.}.\\

We begin in Sec. \ref{sub_gaussianSFT} with a discussion of the
`asymptotically Gaussian' statistics of the SFF, in the
$D\to \infty$ limit. On this basis, we will discuss the size of the fluctuations
in empirical, i.e., numerical averages, explaining the first and third term in
Eq.~\eqref{eq_main}. In Sec.~\ref{sub_sign-kernel}, we will then employ  the
sine-kernel formalism to derive the subleading corrections to the Gaussian
result, namely the second term in Eq. \eqref{eq_main}.\\

\subsection{Gaussian statistics of the SFF for $D\to \infty$}
\label{sub_gaussianSFT}
We begin by formulating a simple argument for the approximately Gaussian
behavior. Let us define
the random variable
\begin{equation}
    u_t = \tr (U_t) 
    = \sum_{i}e^{i \lambda_i t} 
    \ ,
\end{equation}
and represent it as the Fourier transform of the density of states 
\begin{equation}
    \label{eq_u_t}
    u_t = \int \! d\lambda \, \rho(\lambda) ~ e^{i\lambda t}  \ ,
\end{equation} 
In the CUE, the constancy of $\bar\rho$ implies the vanishing of its average $\overline u_t$ for any $t>0$, namely
\begin{equation}
    \label{eq_ut_CUE}
\overline{u_t}= \int \! d\lambda \, \overline \rho(\lambda) ~ e^{i\lambda t}  = \frac {D}{2\pi} \delta(t)\ .
\end{equation}
By contrast, the Fourier transform of the semi-circular spectral density of the GUE is given by
\begin{equation}
    \label{eq_ut_GUE}
    \overline{u_t} 
=  \int \! d\lambda  \, \overline \rho(\lambda) ~ e^{i\lambda t}
= D\frac{J_1(2t)}{t}\ ,
\end{equation}
where $J_1(t)$ is the Bessel function of the first kind. For times
$t\gtrsim 1$ exceeding the microscopically short inverse of the band width, this
function, too, is approximately vanishing. As discussed in the introduction, we refer at the time regime where $\bar u_t$ is suppressed as the ``ergodic regime''.
\\
Hence, the fluctuation part
\begin{equation}
    u_f(t) = u_t - \overline{u_t},
\end{equation}
is the Fourier Transform of $\rho_f(\lambda) =
\rho(\lambda)-\overline{\rho}(\lambda)$. In the ensembles under consideration,
in the large $D$ limit, the spectral density becomes approximately  Gaussian
distributed \cite{haakeSommers1999, cipolloni2023spectral} with variance 
\[\bar \rho_c(\lambda_1, \lambda_2)\equiv \overline{\rho(\lambda_1) \rho(\lambda_2)}-\overline \rho(\lambda_1) \overline \rho(\lambda_2).\]
This statistics is inherited by $u_t$, implying that  in the large $D$
limit $u_f(t)$ is a Gaussian random variable with zero average and variance 
\begin{equation}
    \overline{|u_f|^2}= D z(t)\ ,
\end{equation} 
where $z(t)$ -- defined in Eq. \eqref{eq_zt} -- encodes universal properties of
the spectrum and can be computed via RMT techniques as discussed below in Sec.
\ref{sub_sign-kernel}. On this basis, the moments of the SFF, which are the
moments of $u_f(t)$,  are those of a Gaussian random variable \footnote{The
moments of a Gaussian complex variable $u$ with zero average and variance
$\sigma^2$ read $\overline{|u|^{2n}}=n! \, \sigma^{2\,n}$.}:
\begin{equation}
    \label{eq_gauss_moment}
    \lim_{D\to \infty }\overline{ \SFF_c^n} = \lim_{D\to \infty }\overline{|u_f(t)|^{2n}} = n! \, [D z(t)]^n.
\end{equation}

\begin{figure}[t]
	\centering
\includegraphics[width=1 \linewidth]{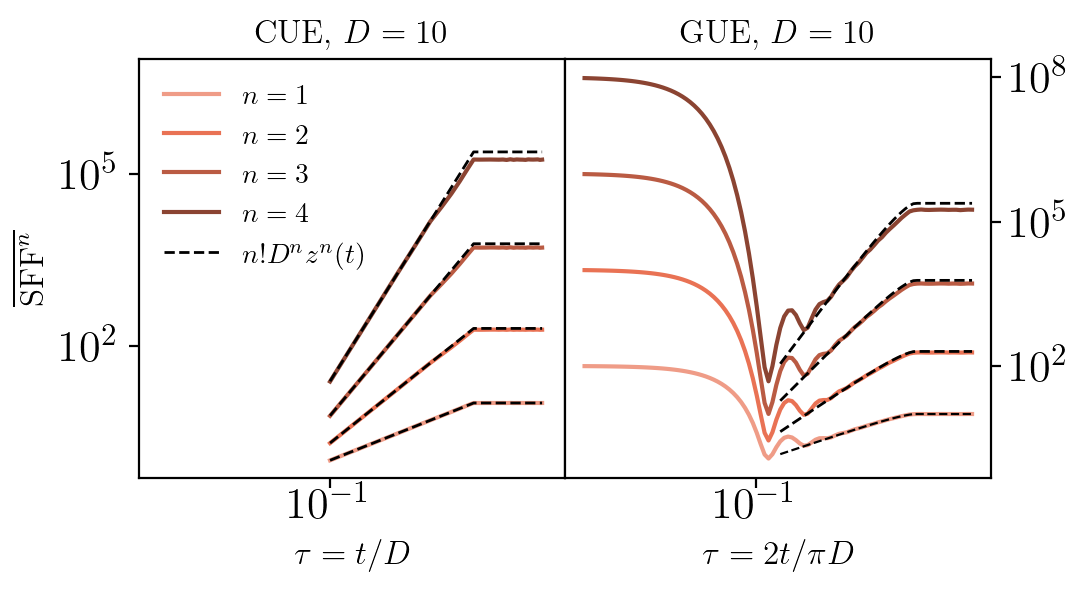}
\caption{Higher-order moments of the spectral form factor, $\overline{\SFF^n}$
for $n=1,2,3,4$ (orange lines with increasing intensity) compared with the
Gaussian prediction, cf. Eq. \eqref{eq_gauss_moment} (black dashed line).
Results are sampled from the CUE (left panel) and for the evolution of a GUE
Hamiltonian (right panel) using $z(t)$ from Eq. \eqref{eq_haake} and Eq.
\eqref{eq_brezin} in the two panels. The numerical data at fixed $D=10$ averaged
over $N_{\rm sim} = 5\cdot 10^5$ samples.}
	\label{fig_3}
 \vspace{-.5cm}
\end{figure}

This behavior is confirmed by numerically evaluating the moments of the spectral
form factor up to $n=4$, as reported in Fig.~\ref{fig_3} in the two ensembles of
interest. As the plot shows, the numerical averages of $\overline{\SFF^n}$
follow the Gaussian prediction $n! D^n z^n(t)$, with some deviations, which are
visible here for a small matrix of size $D=10$.

\subsubsection{Variances and empirical averages}
\label{subsub_variance}
Unlike many other observables characterizing random systems, the SFF is not
self-averaging in the $D\to\infty$ limit \cite{prange1997spectral,
braun2015self}. Rather, its approximate Gaussianity in the large $D$ limit,
implies fluctuations of about the same size as the average: 
\begin{equation}
    \begin{split}
   {\rm Var}(\SFF) & = \overline{\SFF^2}-\overline{\SFF}^2 
   \\ & = [D z(t)]^2 
   = \overline{\SFF}^2\ ,      
    \end{split}
\end{equation}
as numerically verified in Fig.\ref{fig_0}.
This property is true at every order $n$: via Eq. \eqref{eq_gauss_moment}, one has
\begin{equation}
    \label{eq_varia}
    \begin{split}
   {\rm Var}(\SFF^n) & =  \overline{\SFF^{2n}}-\overline{\SFF^n}^2 \\
   & = ((2n)! - (n!)^2)[D z(t)]^{2n}         \propto \overline { \SFF^n}^2\ ,
    \end{split}
\end{equation}
showing that the variance of the $n$th moment scales as the square of its average.

This property has implications for the empirical (i.e., numerical) computation of expectation values
$\overline{[\, \cdot \,]}_{N_{\rm sim}}$ averaged over $N_{\rm sim}$ samples. According 
to the central limit theorem, the fluctuations are suppressed with $N_{\rm
sim}^{-1/2}$ with an amplitude proportional to the standard deviation, i.e.
\begin{equation}
\label{eq_scalingoscilla}
\begin{split}
    \lim_{D\to \infty}\overline{\,[ \SFF^n]\,}_{N_{\rm sim}} & = \overline{\, \SFF^n\,} + \mathcal O(\sqrt{\rm {Var}(\SFF^n)}\,  N_{\rm sim}^{-1/2})    
    \\
    & = D^n \left (n! z^n(t) + \mathcal O(z^n(t)  N_{\rm sim}^{-1/2}) \right )\ ,
\end{split}
\end{equation}
where in the second line, we have used Eq. \eqref{eq_varia} and
Eq. \eqref{eq_gauss_moment}. We thus have  recovered the first and third terms in
Eq. \eqref{eq_main}. This shows that the sampling fluctuations vanish as the
number of samples $N_{\rm sim}$ goes to infinity, making the average
well-defined. Note, however, that these oscillations may be relevant when
addressing numerically subleading contributions $\propto D^{-1}$ and, in those cases, one shall choose
$N_{\rm sim}$ adequately large.

\begin{figure}[t]
	\centering
\includegraphics[width=1 \linewidth]{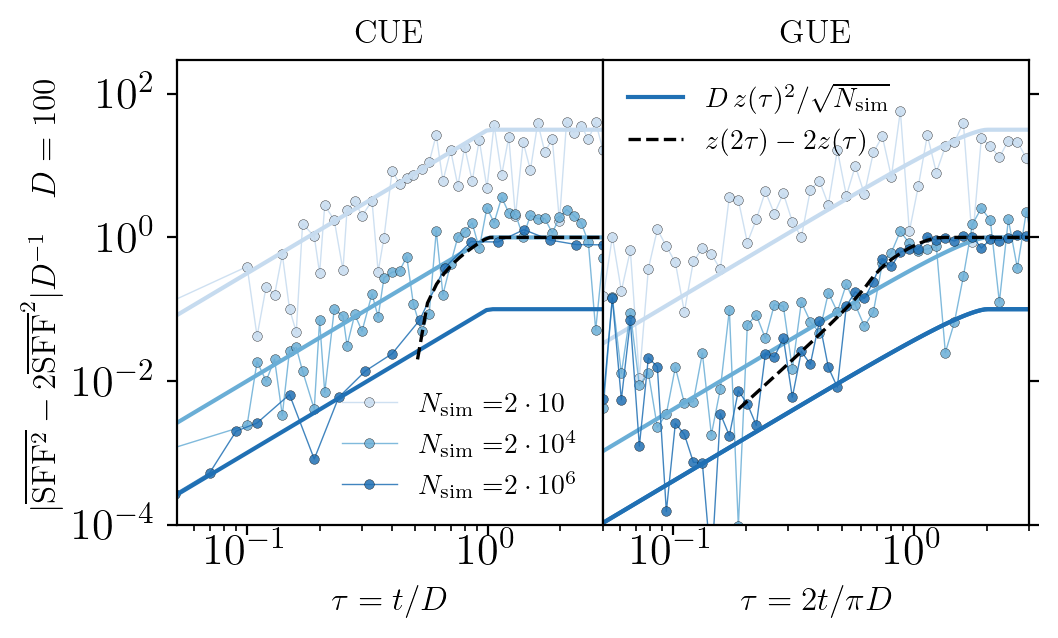}
\caption{Oscillations in the subleading correction to the second moment of the spectral form factor for different empirical averages over $N_{\rm sim}$ simulations. Results are sampled from the CUE (left panel) and for the evolution of a GUE Hamiltonian (right panel). The numerical data at fixed $D=100$ averaged over $N_{\rm sim}=20, 2\cdot 10^4, 2\cdot 10^6$ (blue lines with increasing color intensity) are compared with the prediction in Eq. \eqref{eq_main}, with full blue lines we indicate the amplitude of the oscillatory term $\mathcal O(D z^2(\tau)/\sqrt{N_{\rm sim}})$ and with a dashed black line the subleading correction in Eq. \eqref{eq_conn}, using Eq. \eqref{eq_haake} and Eq. \eqref{eq_brezin} in the two panels. }
	\label{fig_4}
 \vspace{-.5cm}
\end{figure}

Fig.~\ref{fig_4} shows  data for $\frac{1}{D}|\overline{\SFF^2}-2\overline{\SFF}^2|$
with $D=100$. For few simulation runs, the distribution of data points reflects the fluctuations due to the
finiteness of  $N_{\rm sim}$, and it is centered around the estimate functions
$D z^2(t)/N_{\rm sim}^{-1/2}$, indicated by solid blue lines. However,  for
large values $N_{\rm sim}\gg D^2$, the fluctuations   converge to a different
function, namely the genuine deviation from Gaussianity $z(2t)-2z(t)$ defined in
Eqs.~\eqref{eq_main} and \eqref{eq_conn}, and represented by a dashed black line
in the figure. The goal of the next section is to understand the origin of this
subleading term using the tools of RMT.

\subsection{Sine-Kernel predictions}
\label{sub_sign-kernel}

In this section, we review the random matrix theory framework for analyzing
moments of the SFF. We closely follow Ref. \cite{Liu2019spectral}, where
explicit calculations are presented at a high level of detail. However, we also
correct a number of inconsistencies which ultimately led to incorrect final
results in that reference.

Before delving into the analysis of the SFF, let us review a few elements of RMT
spectral statistics. Random matrix theory  describes the energy spectra  in
terms of  correlation functions defined as 
\begin{align}
    \label{eqRnDef}
   &R_n(\lambda_1, \ldots, \lambda_n) =
\sum_{i_1 \neq i_2 \neq \dots\neq i_n } \! 
\overline
{\delta(\lambda_1-\lambda_{i_1})
\dots \delta(\lambda_n-\lambda_{i_n})
}\ ,
\end{align}
where the overbar denotes averaging over the statistical ensemble of the eigenvalues of either of
Gaussian or Haar distributed random operators (the results summarized in this
section apply to both cases unless explicitly noted). These functions are
commonly defined to exclude the auto-correlation of identical levels, as
indicated by the condition $i_1\neq i_2\dots$ beneath the sum. This exclusion
leads to modifications relative to the unconditioned sum, which are small in
$D^{-1}$, but become important when it comes to higher order cumulative
correlations --- we need to keep an eye on them \footnote{We note that Refs.\cite{cotler2017chaos, Liu2019spectral}, work with the re-scaled $n$-point correlation functions 
\begin{equation}
    \rho^{(n)}(\lambda_1, \dots, \lambda_n) = \frac{(D-n)!}{D!}R_n(\lambda_1, \dots, \lambda_n)\ ,
\end{equation}
 defined such that the average is normalized to one, i.e. $\int d\lambda \rho^{(1)}(\lambda)=1$.}.

The two lowest-order correlation functions of this type are 
\begin{equation} 
    R_1(\lambda) 
    = \overline{\rho}(\lambda), 
\end{equation}
defining the average of the spectral density 
\begin{align}
    \label{eq:RhoDef}
    \rho(\lambda)=\sum_{i}\delta(\lambda-\lambda_i),
\end{align}
normalized as $\int \! d\lambda  \, R_1(\lambda)=\int \! d\lambda \, \overline{\rho}(\lambda)  =D$,
and 
\begin{equation} 
    \label{R2_def}
    R_2(\lambda_1, \lambda_2) = \overline{\rho(\lambda_1)\rho(\lambda_2)} - \delta(\lambda_1-\lambda_2) \overline{\rho}(\lambda_1), 
\end{equation}
where the subtraction removes the auto-correlation part entering the first term. 

A crucial result of RMT expresses the correlations between $n$ levels as the determinant of an $n$-dimensional matrix,    
\begin{equation}
    R_n(\lambda_1, \ldots, \lambda_n) =
    \text{det}\left(K_D(\lambda_i, \lambda_j)\right)_{i,j=1}^{n},
    \label{eq:rho_n}
\end{equation}
where the specific form of the kernel $K_D(\lambda,\lambda')$ depends on the
symmetries of the matrix ensemble, the magnitude of the spacings
$|\lambda-\lambda'|$, and on the distinction unitary vs. hermitian matrix
realization. However, in the case $|\lambda-\lambda'|\sim D^{-1}$, where eigenvalues are close on the scale of the level spacing, a limit of high universality is realized.
The kernel $K_D(\lambda,\lambda')=K_D(\lambda-\lambda')$ then depends only on
eigenvalue differences, with a single-argument function $K_D(\lambda)$ solely determined by
symmetries. Specifically, for class A (aka GUE/CUE) describing hermitian or
unitary random operators, excluding further symmetry constraints, $K_D$ is the
celebrated sine kernel defined as
\begin{equation}
    \lim_{D\to \infty}K_D(\lambda_i, \lambda_j) = 
    \begin{dcases}
        \overline{\rho}\,  \frac{\sin \left(\pi \overline \rho (\lambda_i - \lambda_j)\right)}{\pi \overline \rho(\lambda_i - \lambda_j)} & \text{for } i \neq j, \\
        \bar \rho & \text{for } i = j,
    \end{dcases}\ 
    \label{eq:kernel_K}
\end{equation} 
where $\bar \rho\approx \bar \rho(\lambda_{i})\approx \bar \rho(\lambda_{j})$ is the
characteristic spectral density at the point where the spectrum is probed. 

On this basis, we now turn to the analysis of the  spectral form factor, whose $k$th moment is defined as
\begin{align} 
\label{eq_sffn}
\begin{split} 
\overline{\text{SFF}^{k}} &= \overline{\left(\Tr(e^{-iHt})\Tr(e^{iHt})\right)^k}\ 
\\
& =\overline{\sum_{i_1 \dots i_k,j_1 \dots j_k} \! e^{i(\lambda_{i_1}+\dots+\lambda_{i_k}-\lambda_{j_1}-\dots-\lambda_{j_k})t}} \\
&=\int_\lambda \, \overline{\rho(\lambda_1)\dots \rho(\lambda_{2k})}~e^{i(\lambda_1 + \dots + \lambda_k - \lambda_{k+1}-\dots -\lambda_{2k}) } 
\end{split} 
\end{align}
in terms of eigenvalues, where here and throughout $
\int_\lambda$ is shorthand notation for the integration over all $\lambda$-variables. In the following, we use this representation as a starting point to obtain explicit representations in terms of random matrix correlation functions for the cases $k=1$ and $2$. 

\subsubsection{First moment of the SFF}

As a warmup, consider Eq.~\eqref{eq_sffn} for $k=1$, 
\begin{align}
 \overline{\SFF}= \int_\lambda  
 \overline{\rho(\lambda_1)\rho(\lambda_2)} ~ e^{i(\lambda_1-\lambda_2)t}.
\end{align}
Using the definition of $R_2$ in Eq. \eqref{R2_def} and its expression given by
determinant formula in Eq. \eqref{eq:rho_n}, i.e.
\begin{align*}
    R_2(\lambda_1, \lambda_2) =
\overline{\rho}(\lambda_1)\overline{\rho}(\lambda_2)-K(\lambda_1, \lambda_2)^2,
\end{align*}
we obtain 
\begin{align}
\begin{split}
   \overline{\SFF} & =D+\int_\lambda  R_2(\lambda_1,\lambda_2) ~ e^{i(\lambda_1-\lambda_2)t} 
   \\
   & = 
    D + |\overline{u_t}|^2- \int_\lambda K_D(\lambda_1, \lambda_2)^2 ~ e^{i(\lambda_1-\lambda_2)t}
   \end{split}
\end{align}
where $\bar u_t$ is the Fourier transform of the average density of states, cf. Eqs.~(\ref{eq_u_t}), (\ref{eq_ut_CUE}), (\ref{eq_ut_GUE}). Accordingly, the connected part of the SFF (cf. Eq.~\eqref{eq_main}), 
    $\overline{\SFF_c} = \overline{\SFF}-|\overline{u_t}|^2$,
is determined by the sine-kernel as \[\overline{\SFF_c}=D z(t),\] with
\begin{align}
\begin{split}
    z(t) & 
    = 1 - \frac 1D \int_\lambda  K_D(\lambda_1, \lambda_2)^2 ~e^{i(\lambda_1-\lambda_2)t}    \ .
    \label{eq_z_RMT}
\end{split}
\end{align}
Assuming a constant average density of states $\bar \rho \equiv t_\textrm{H}/2\pi$, the Fourier transform of  Eq.~\eqref{eq:kernel_K} 
then leads to the celebrated result
\begin{align}
    \label{eq_rampCUE}
     \overline{\SFF_c}=
     D~\begin{dcases}
       \tau & \text{for } \tau < 1, \\
        1 & \text{for }  \tau>1,
    \end{dcases} \qquad \tau\equiv \frac{t}{t_\textrm{H}},
\end{align}
 showing the characteristic ramp and plateau structure with linear growth followed by instantaneous saturation at $t=t_\textrm{H}$. 

For the GUE,  the
non-constancy of the average spectral density leads to modifications of this
result. A detailed analysis of the integral in Eq. \eqref{eq_z_RMT} with account
for an energy-dependent $\bar \rho(\lambda)$  in Eq. \eqref{eq:kernel_K} yields
Eq. \eqref{eq_brezin} \cite{brezin1997spectral}, which provides an improved basis for the comparison with
numerical simulations for values of $D$ where these modifications become sizable.

\subsubsection{Second moment of the SFF}
Turning to the second moment, $k=2$ in Eq.~\eqref{eq_sffn}, we start from the formal eigenvalue sum
\begin{align}
    \label{eqSFFEigenvalues}
   \overline{\SFF_2} =\sum_{i_1,i_2,j_1,j_2} \overline{e^{i(\lambda_{i_1}+ \lambda_{i_2}- \lambda_{j_1}-\lambda_{j_2})t}}.
\end{align}
The first task now is to organize this sum into contributions with a definite
number of pairwise equal indices. These terms can then be individually
represented in terms of the correlation functions $R_{1,\dots,4}$,
Eq.~\eqref{eqRnDef}, leading to
\begin{align}
    \begin{split}
       \overline{\SFF^2}=& 
       \int_\lambda R_4(\lambda_1,\lambda_2, \lambda_3, \lambda_4) ~e^{i(\lambda_1 + \lambda_2 - \lambda_3 - \lambda_4)t}  \\
    & + 2
    \int_\lambda  R_3(\lambda_1,\lambda_2, \lambda_3) ~e^{i(2\lambda_1 - \lambda_2 - \lambda_3 )t} 
    \\
    & + \int_\lambda R_2(\lambda_1,\lambda_2) ~e^{2i(\lambda_1 - \lambda_2)t} \\
    & + 4(D - 1)\int_\lambda R_2(\lambda_1,\lambda_2) ~e^{i(\lambda_1 - \lambda_2)t}  \\
    & + 2D^2 - D.  
    \label{eq:R4_total}
    \end{split}
\end{align}
Here, the third and the fourth line have already been computed in terms of $z(t)$ in Eq. \eqref{eq_z_RMT} as
\begin{align*}
    \int_\lambda R_2(\lambda_1,\lambda_2) ~ e^{i(\lambda_1 - \lambda_2)t} = D[z(t)-1] + |\overline{u_t}|^2\ .
\end{align*}
Turning to the first and second line, we use Eq.~\eqref{eq:rho_n} to represent
them through determinants of rank 4 and 3 matrices, and then do the integral
over energy arguments. Referring to Appendix~\ref{app_sinekernel} for details,
this leads to 
\begin{subequations}
\begin{align}
\begin{split}
    \int_\lambda & R_3(\lambda_1,\lambda_2, \lambda_3) ~e^{i(2\lambda_1 - \lambda_2-\lambda_3)t} \\ & = 2D \left [1-z(2t) \right ]
    + \Re (\overline{u_{2t} }\, \overline{u_{-t}}^2) \ ,   
    \label{eq: R_3}
\end{split}
\end{align}
\begin{align}
\begin{split}
    \int_\lambda & R_4(\lambda_1,\lambda_2, \lambda_3, \lambda_4)~e^{i(\lambda_1 + \lambda_2-\lambda_3-\lambda_4)t} \\
    & = 2D^2 [1-z(t)]^2 - 2 D[3-z(t) -2 z(2t )]\\
    & \quad + |\overline{u_t}|^4-4D |\overline{u_t}|^2
     [1-z(t)] \ .
    \label{eq: R_4}
\end{split}
\end{align}    
\end{subequations}
Summing over all the contributions, the second moment of the SFF reads
\begin{align}
   \overline{ \SFF^2} = & |\overline{u_t}|^4 +  [\, |\overline{u_{2t}}|^2 - 4 |\overline{u_t}|^2 \,]\nonumber \\
          & + 2[ \Re(\overline{u_{2t} }\, \overline{u_{-t}}^2) + 2 |\overline{u_t}|^2 D z(t)] \nonumber \\
          & + 2D^2 z(t)^2 + D \left [z(2t)-2z(t)\right] \ ,
\label{eq:R4_final_result}
\end{align}
Thus, neglecting the rapidly vanishing factors $\overline{u_t}$, this equation simplifies to
\begin{align}
    \overline{\SFF^2} \simeq  2D^2 z(t)^2 + D \left[z(2t)-2z(t)\right] \ .
    \label{eq:R4_approximate}
\end{align}
where the first term corresponds to the asymptotic Gaussian result discussed in
Sec.~\ref{sub_gaussianSFT}.  However, the second term defines the sub-leading
non-Gaussian contribution in Eq. \eqref{eq_main} as
\begin{equation}
    {\rm Conn}_2(t) = z(2t)-2z(t)\ .
    \label{eq_conn2}
\end{equation}
For the CUE, where  $z(t)$ is given by Eq.(\ref{eq_haake}), it assumes the form
\begin{align}
\label{eq_conn2_RMT}
     {\rm Conn}_2(t) =
     \begin{cases}
       0  & \text{for } t < \frac{t_{\rm H}}{2}, \\
        1-2t/t_{\rm H} & \text{for }  \frac{t_{\rm H}}{2}<t<t_{\rm H},\\
        -1    &         \text{for } t>t_{\rm H}.
    \end{cases}
\end{align}
Adding the Gaussian contribution, we obtain the final result
\begin{align}
    \label{eqSFFSecondMoment}
     \overline{\SFF^2} = D^2
     \begin{dcases}
       2 \tau^2  & \text{for } \tau < \frac{1}{2} \\
        2 \tau^2+(1-2\tau)/D & \text{for }  \frac{1}{2}<\tau<1\\
        2 - 1/D    &         \text{for } \tau>1 
    \end{dcases}\ ,
\end{align}
first obtained in \cite{haakeSommers1999} by a different method operating entirely in the time domain.  
Eq. \eqref{eqSFFSecondMoment} describes the second moment of the form factor,
distinguishing between three time domains. The quadratic growth for early times
$(t<t_{\rm H}/2)$ represents purely Gaussian statistics, i.e.
$\overline{\SFF^2}=(\overline{\SFF})^2+ \textrm{var}(\SFF)$, with
$\textrm{var}(\SFF)=\overline{\SFF}^2=\tau^2$. This is a regime accessible to
perturbation theory, as discussed in Sec.~\ref{sec_gauss_ft}. The constancy for late times,
$\tau>1$, can be heuristically understood from Eq.~\eqref{eqSFFEigenvalues}. In
this regime, only contributions without oscillatory phases survive. Elementary
combinatorics shows that there are $2D^2-D$ of these, explaining the constant.
If the perturbative result $\sim \tau^2$ would extend up to $\tau=1$, we would
observe a mismatch between $2D^2$ (perturbative) and $2D^2-D$ (plateau). The form factor
avoids this problem by sandwiching in a third regime between $\tau=1/2$ and
$\tau=1$. The sudden (non-analytic) change at $\tau=1/2$ hints at the
non-perturbative nature of this intermediate regime.

Finally, for the  GUE
ensemble, the substitution of Eq.~\eqref{eq_brezin} into Eq.~\eqref{eq_conn2} leads to a refined representation which we do not display explicitly but use in our comparison with simulations. 

This result is in agreement with numerics, as shown in Fig.~\ref{fig_1} above
for both the CUE and GUE. 

The sine-kernel approach naturally generalizes to the computation of higher-order moments $\overline{\textrm{SFF}^k}$. The method remains structurally similar, with the $k$-th moment involving $2k$-point spectral correlation functions, expressible as determinants of $2k\cross 2k$ sine-kernel matrices. While the analytic complexity increases with $k$, the structure remains governed by universal kernel identities. As an illustration, for $k=3$, the dominant Gaussian contribution is $\overline{\mathrm{SFF}^3} \approx 6 D^3 z(t)^3$, while the leading subleading correction arises from connected contractions in the six-point correlation function $R_6$ and is expected to involve terms such as $z(3t)$, $z(2t)z(t)$, and their combinations. Although we do not compute this correction explicitly, the method's extensibility is evident, and such terms can be systematically extracted by extending the combinatorics developed similarly to  Eq.(\ref{eq:R4_total}) and computing the various integrals, such as
    $\int_{\lambda} R_k(\lambda_1, \lambda_2, \dots \lambda_k)e^{i \lambda_1 t_1 + i \lambda_2 t_2 + \dots + i \lambda_k t_k}$.

In the Section below, we show how to retrieve the same
behavior using field theory computations, reinforcing the validity of our
corrected approach.

\section{Field-theory calculations}
\label{sec_fieldtheory}

In this Section, we will discuss the computation of SFF statistics by  field
theory techniques. 
However, before turning to the discussion of the SFF in terms of such methods in
Section \ref{sec_non_pertuba}, it is worth taking a look at the situation within
the framework of perturbation theory in $D^{-1}$. The advantage of this
approach, reviewed in Sec. \ref{PerturbationTheory}, is that it keeps
transparent track of this expansion parameter at all stages of the computation and
provides an intuitive explanation for the Gaussian correlation of the SFF
discussed in Section \ref{sub_gaussianSFT}. Readers primarily interested in
final results may skip the next section and proceed directly to Section
\ref{sec_non_pertuba}.
 
\subsection{Asymptotic Gaussian SFF 
 from perturbation theory}
\label{PerturbationTheory}
For definiteness, we consider a GUE ensemble of Hamiltonians $H=\{H_{\mu\nu}\}$ distributed
according to Eq.~\eqref{eq:GaussianStatistics} with second moment $\overline{ |H_{\mu\nu}|^2} =D^{-1}$.
The perturbative approach is best formulated in energy-Fourier space, where the
retarded Green function, or resolvent,
\begin{align}
    \label{eq_resolvent}
    G^+(\lambda)\equiv -i\int_0^\infty \mkern-15mu dt ~ U_t ~e^{i(\lambda+i0) t}=(\lambda^+-H)^{-1}
\end{align}
and its advanced sibling are center stage, where we use the short-hand notations $\lambda^\pm=\lambda \pm i 0$. An expansion of these objects in $H$
yields formal series, which we need to analyze to leading order in a $D^{-1}$
expansion. This analysis is best organized in a two-stage process, where one
first computes $\bar G^\pm \equiv \overline{ G^\pm }$, before turning to the computation of
correlations between different Green functions.  
\label{sec_gauss_ft}

\subsubsection{Average Green function}

\begin{figure}[h]
	\centering
\includegraphics[width=0.9 \linewidth]{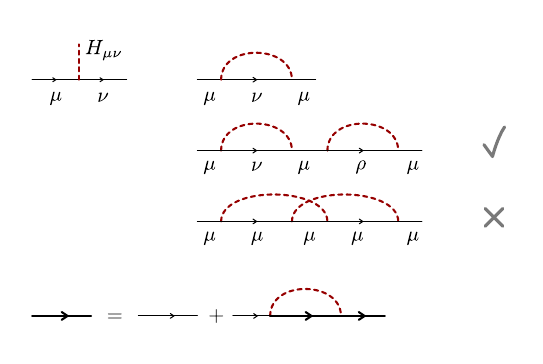}
\caption{Top left: graphical representation of the expansion of the resolvent in Eq.~\eqref{eq_resolvent} to
first order in $H$. Top right: The averaging of a second order term leads to
index correlations as indicated. Right center: To leading order in the $D^{-1}$
expansion, each additional order in $H^{2}$ must be accompanied by a free
running index summation. The diagram shown satisfies this criterion.
Right bottom: a suppressed diagram, which we disregard. Bottom: the recursive
summation of all diagrams without crossing lines defines the SCBA.}
	\label{fig:SCBA}
 \vspace{-.25cm}
\end{figure}

The way in which the parameter $D\gg 1$ facilitates the computation of the Green
function is illustrated in Fig.~\ref{fig:SCBA}. At each additional order of the expansion of the resolvent in Eq.~\eqref{eq_resolvent} in
even powers of $H$, we must pick up a free running summation over $\mu$ to
compensate for the $D^{-1}$ in the GUE variance. In the diagrammatic code used
in the figure, this amounts to the exclusion of the diagrams with intersecting
lines. The survivor class of non-crossing diagrams can be summed up, and leads
to the equation 
\begin{align*}
    \bar G^\pm(\lambda)= \frac{1}{\lambda^\pm - \bar G^\pm(\lambda)}.
\end{align*}

Alluding to the resemblance of the denominator to that of a quantum state
    subject to a `self energy' $\Sigma = G$, this equation is known as the \emph{self-consistent Born approximation} (SCBA), or Pastur equation in the context of Random Matrix Theory \cite{pastur1973spectra}. The
straightforward solution of the quadratic equation leads to 
\begin{align}
    \label{eq:BarGPertTheory}
    \bar G_{\mu \nu}^\pm(\lambda) = \frac{\delta_{\mu\nu}}{\frac{\lambda}{2}\pm i \Sigma(\lambda)}= \delta_{\mu \nu}\left( \frac{\lambda}{2}\mp i \Sigma(\lambda) \right),
\end{align}
with 
\begin{align*}
    \Sigma(\lambda)=\left(1-\left(\frac{\lambda}{2}\right)^2\right)^{1/2}.
\end{align*}
From here, the average spectral density is seen to realize the celebrated
semi-circular profile, 
\begin{align*}
    \overline \rho(\lambda)=-\frac{1}{2\pi i}{\rm tr}(\bar G^+(\lambda)-\bar G^-(\lambda))=\frac{D}{\pi}\Sigma(\lambda).
\end{align*}

\subsubsection{Average SFF}
\begin{figure}[h]
	\centering
\includegraphics[width=0.9 \linewidth]{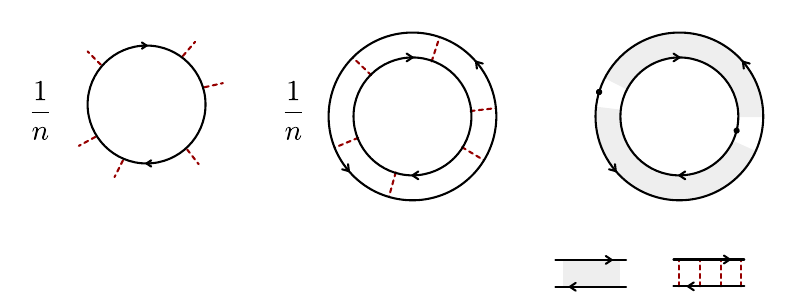}
\caption{Left: perturbative expansion of the Green function generating function. The logarithmic form of the generating function leads to a factor $1/n$ at $n$th order. Center: The average of two generating functions, where the direction of the arrows distinguishes between retarded and advanced resolvent. Notice the thickened lines indicating that the individual resolvents are SCBA dressed. Right: diagrammatic interpretation of the correlation function between two closed loops (traces) of Green functions. The shaded regions represent ladder diagrams, i.e. geometric series of non-crossing scattering processes. This can be interpreted as a heuristic version of the resolvent method in Ref.~\cite{cipolloni2023spectral}.}
	\label{fig:SFFGeneratingFunction}
 \vspace{-.5cm}
\end{figure}

With these results in place, we can readily turn to the computation of the
average SFF. Working in the energy representation, we represent the spectral
density as the difference between retarded and advanced Green functions, to obtain
the connected correlation function  $\bar \rho_c(\lambda_1,\lambda_2)$
featuring in Eq.~\eqref{eq_zt} as 
\begin{align*}
    &
    \bar \rho_c(\lambda_1,\lambda_2)
    = -\frac{1}{(2\pi)^2} \sum_{s_1,s_2=\pm} s_1 s_2
    X^{s_1 s_2},\\
    &\qquad X^{s_1 s_2}\equiv \overline{{\rm tr}(G^{s_1}(\lambda_1))\, 
    {\rm tr}(G^{s_2}(\lambda_2))}. 
\end{align*}
Notice that $X$ here and in the following only accounts for the connected part of the correlation function.
Let us first focus on the building block $X^{+-}$. An economical way to compute
it in perturbation theory represents the traces as derivatives of `generating
functions': $\tr(G^\pm(\lambda))=\partial_{\lambda^\pm}\tr\ln(\lambda^\pm-H)\ .$
Now, consider the representation
\begin{align*}
    X^{+-}(\lambda_1,\lambda_2)=\partial^2_{\lambda_1
    \lambda_2}\overline{ \tr\ln(\lambda_1^+-H)\tr\ln(\lambda^-_2-H) }
\end{align*}
expanded
in a formal series in $H$. To leading order in the $D^{-1}$ expansion, the
subsequent average over $H$ will then select all diagrams excluding
intersections, and the corresponding contraction lines fall into two categories,
lines connecting individual logarithmic factors with themselves, and lines
between the two logarithms. The summation over the former amounts to a
replacement of the resolvent operators featuring as arguments of the $\tr\ln$ by
the dressed resolvents discussed in the foregoing section. For a visualization
of the ladder structures connecting the two logarithms, we refer to Fig.~\ref{fig:SFFGeneratingFunction}. Focusing on a region where $|\lambda_1-\lambda_2|\sim D^{-1}$, one can expand the density of state around the average energy $\lambda\equiv (\lambda_1+\lambda_2)/2$, and the resummation of the resulting diagrams leads to the result
\begin{align*}
    X^{+-}(\lambda_1,\lambda_2)& \simeq \partial^2_{\lambda_1,\lambda_2}\ln\left( 1-\frac{1}{D}{\rm tr}\left(\bar G^+(\lambda_1)\bar G^-(\lambda_2) \right)\right)\\
    &\simeq \partial^2_{\lambda_1,\lambda_2}\ln\left( \frac{i\pi \bar \rho(\lambda)}{2D}(\lambda_1-\lambda_2) \right) \\
    &\simeq -\frac{1}{(\lambda_1-\lambda_2)^2}\ ,
\end{align*}
where the ``$\simeq$'' are equalities up to corrections of subleading order in
$D^{-1}$. For an interpretation of the two-fold derivative in the language of
diagrams, we again refer to Fig.~\ref{fig:SFFGeneratingFunction}. Turning to the
remaining terms $X^{ss'}$, $X^{+-}$ is symmetric in $\lambda_1$ and $\lambda_2$, implying that
$X^{-+}(\lambda_1,\lambda_2)=X^{+-}(\lambda_1,\lambda_2)$. However, the
uni-causal terms $X^{++}$ and $X^{--}$ contain the essential logarithm in the
form $\ln\left(1- \frac{1}{D}{\rm tr}\left(\bar G^s \bar G^s\right)\right)$. Since $\bar G^s \bar G^s \approx -1$, we
obtain no singularity $\lambda_1-\lambda_2$.  Upon Fourier transformation back
to time, these terms generate short-lived contributions $\sim \delta(t)$,
where the $\delta$-function is smeared over scales $\sim \lambda^{-1}$. We
neglect them as inessential in the finite time form factor. Adding the
contributions $X^{+-}+X^{-+}$, we obtain the perturbative result 
\begin{align}
    \label{eq_saddle}
    \bar\rho_c(\lambda_1,\lambda_2)\approx -\frac{1}{2}\left( \frac{1}{\pi(\lambda_1-\lambda_2)} \right)^2
\end{align}
which, when substituted back into Eq.~\eqref{eq_zt} and subsequently Fourier transformed, leads to the
linear ramp of the SFF. 

\subsubsection{Second Moment of SFF}
\begin{figure}[h]
	\centering
\includegraphics[width=0.7 \linewidth]{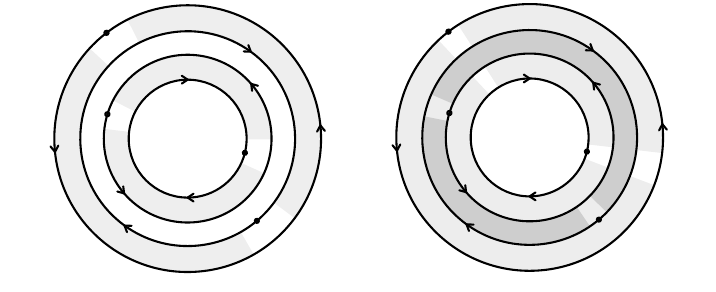}
\caption{To leading order in $D^{-1}$, i.e. to an order satisfying the
one-summation-per-scattering-process rule, products of $2n$ generating functions
statistically decouple, as indicated here for $n=2$. The sum of the two diagrams
on the right-hand-side defines the second moment of the SFF. Notice that the
individual propagators (distinguished by shading for improved visibility)
correspond to energy denominators
$1/(\lambda_{1+}-\lambda_{1-})^2(\lambda_{2+}-\lambda_{2-})^2$ and
$1/(\lambda_{1+}-\lambda_{2-})^2(\lambda_{2+}-\lambda_{1-})^2$, respectively.} 
	\label{fig:SFFSecondMoment}
 \vspace{-.5cm}
\end{figure}

The framework above affords straightforward generalization to the $n$th moment of
the SFF. All we need to do is consider $2n$ generating function factors with
appropriately multiplied energy arguments. It is now straightforward to convince
oneself that to leading order in the $D^{-1}$ expansion there are no 
diagrams coupling more than $2$ factors together. In other words, the average
decouples into $n$ factors, as indicated in Fig.\ref{fig:SFFSecondMoment} for the case $n=2$. The
combinatorial factor $n!$ counting the different choices of pairings explains
the approximately Gaussian statistics of the SFF (again, to leading order in $D^{-1}$).   

\subsection{Subleading corrections via the supersymmetric method}
\label{sec_non_pertuba}

In order to access the non-perturbative contributions to the statistics of the
SFF, we employ a supersymmetric approach \cite{efetov_book}. As our focus
will be on universal properties of the deviations from Gaussianity, we consider
the CUE ensemble, where the density of states is constant. We will start from a
representation of SFF moments in terms of ratios of determinants, which are then
expressed as integrals over bosonic and fermionic `ghost' fields. 

The subsequent averaging over Haar distributed randomness is performed by a
variant of the Hubbard-Stratonovich known as color-flavor
transformation~\cite{zirnbauer_color_flavor}. This transformation expresses
moments of the SFF in terms of an
integral over auxiliary variables, which can be performed exactly by saddle
point integration, thanks to a feature known as `semiclassical exactness'. The
results we obtain in this manner are therefore likewise exact. However, before
turning to the execution of this program, some comments are in order. First, the
result for the SFF discussed below is implicit in the previous
Ref.~\cite{conrey2007howepairssupersymmetryratios}, which obtained it by
representation theoretic methods. We here rederive this result in a framework
perhaps more accessible to most physicists. Second, and unlike the method of
Ref.~\cite{conrey2007howepairssupersymmetryratios}, the CFT is known to run into
convergence issues if the number of colors becomes of
$\mathcal{O}(1)$\cite{zirnbauer2021colorflavortransformationrevisited}. At the
same time, numerics for random matrices of the smallest
non-trivial dimensions $D=2,3$ turn out to be in perfect agreement with the
analytical results obtained by CFT, indicating that the saddle point program
works even beyond its range of mathematical backing. Finally, semiclassical
exactness is a precious feature of a few symmetry classes, including class A
discussed in this paper. For instance, the SFF statistics of Haar distributed orthogonal
matrices require a much more elaborate analysis, beyond the scope of this
paper.

\subsubsection{Generating function and non-linear sigma model} 

The supersymmetry approach to the SFF is formulated in a phase
representation, Fourier conjugate to discrete time. We thus start from the
Fourier transform
\begin{align*}
{\rm tr}(U^t)
&=
\int \frac{d\phi}{2\pi}~
e^{-i\phi t}
\,
 {\rm tr}\left(\frac{1}{1-e^{i\phi^+}U}\right),
 \\
 {\rm tr}([U^\dagger]^t)
&=\int \frac{d\phi}{2\pi}~
e^{i\phi t}
\,
 {\rm tr}\left(\frac{1}{1-e^{-i\phi^-}U^\dagger}\right),
\end{align*}
with $t\in \mathbb N$ the discrete time of evolution and $\phi_s^\pm =
\phi_s \pm i 0$ \textbf{}slightly shifted into the upper/lower complex plane. 

This motivates to introduce 
\begin{widetext}
\begin{equation}
\label{eq:gen_function}
    \mathcal{Z}(\alpha, \phi) = \Bigg\langle \frac{\det( 1 - e^{i(\alpha_1 + \phi_1^+)}U^\dagger) \det( 1 - e^{i(\alpha_2 + \phi_2^+)} U^\dagger)\det( 1 - e^{-i(\alpha_3 + \phi_3^-)}U) \det( 1 - e^{-i(\alpha_4 + \phi_4^-)} U)}{\det( 1 - e^{i \phi_1^+}U^\dagger) \det( 1 - e^{i \phi_2^+} U^\dagger)\det( 1 - e^{-i \phi_3^-}U) \det( 1 - e^{-i \phi_4^-} U)} \Bigg\rangle_U,
\end{equation}
\end{widetext} 
as a generating function for the second moments of the form factor.
Indeed, it can be verified that 
\begin{align}
    \overline{\text{SFF}^2} &= \frac{1}{(2\pi)^4} \! \int d\phi ~ {\cal F}(\phi)~  e^{-it(\phi_1+\phi_2-\phi_3-\phi_4)}\\
    {\cal F}(\phi) &= \partial_{\alpha_1} \partial_{\alpha_2} \partial_{\alpha_3} \partial_{\alpha_4} ~\mathcal{Z}(\alpha, \phi) \biggr|_{\alpha=0},
\end{align}
where $\int d\phi$ denotes integration over all $\phi$-variables. The four
determinants in the numerator and denominator of  Eq.~\eqref{eq:gen_function}
then allow for integral-representations involving four complex (`bosonic') and
four Grassmann (`fermionic') variables, respectively. Collecting the eight
integration variables into the `supervector' $\psi$, we obtain the representation
\begin{align}
\mathcal{Z} 
&=   \int_{\mathrm{U}(D)}\mkern-20mu dU \int d(\bar \psi, \psi) ~
e^{ \bar \psi  \left(\mathbf{1}_{8D}- E {\cal U} \right) \psi},
\end{align}where the integration of bosonic variables is along the imaginary axis, and $\mathbf{1}_{8D}$
is the identity matrix operating in the
$8$-fold replicated Hilbert space of 
graded (`bosonic/fermionic'), causal (`retarded/advanced'), and replicated (`1/2') variables. Here, we upgraded the time-evolution operator
\begin{align}
\label{eq:U_replicated}
{\cal U}
&=
\begin{pmatrix} 
U^\dagger & \\ & U \end{pmatrix}_{ra}\otimes \mathbf{1}_2\otimes \mathbf{1}_{bf},
\end{align}
 to act on the replicated Hilbert space, where `flavor' indices `${\rm ra}$', `$2$' and `${\rm bf}$' denote the subspace of causal, replica and graded indices. Furthermore, we introduced   
\begin{align}
\label{eq:def_E}
    E &= \exp(i\Lambda(\phi+\alpha)),
\end{align}
involving sources and phases organized into matrices 
\begin{align}
\label{eq:def_alpha_phi}
    \alpha &= {\rm diag}(\alpha_1, \alpha_2,\alpha_3,\alpha_4)\otimes \pi_{f}, \\
    \phi &= {\rm diag}(\phi_1^+, \phi_2^+,\phi_3^-,\phi_4^-) \otimes \mathbf{1}_{bf},
    \end{align}
where $\pi_f=\frac{1}{2}(1-\sigma^{bf}_3)$ is a projection onto Grassmann components and we omitted the trivial Hilbert-space structure $\mathbf{1}_D$. 

Finally, the matrix 
\begin{align}
\label{eq:Lambda}
 \Lambda 
 &= 
 \sigma_3^{ra} \otimes \mathbf{1}_2 \otimes \mathbf{1}_{bf},
 \end{align}
follows the causal structure of time-evolution Eq.~\eqref{eq:U_replicated} and plays an important role in the following. 

To perform the average over the Haar ensemble we  employ the color-flavor
transformation~\cite{zirnbauer_color_flavor,
haake_quantum_signatures_of_chaos,Review_of_Sigma_Models}, which exchanges the
integral over unitaries $U$, acting on the `color' indices of  Hilbert space, and
structureless in flavor space, for an integral over matrices $Z,\bar{Z}$. These
latter, on the other hand, act  nontrivially on `flavor' indices but are
structureless in `color' space. Referring for   
details to Appendix~\ref{app_field_theory}, we notice that  the transformation followed by $\psi$-integration leads to
\begin{align}
\label{eq:ZZ_sigma_model}
    \mathcal{Z} &=  \int \! d(Z, \tilde{Z})~ e^{D \left(
    \text{Str} \ln(1-\tilde{Z}Z)
    - \text{Str} \ln(1-E_+Z E_-\tilde{Z})
    \right)},
\end{align}
where `${\rm Str}$' is the graded trace and $E_\pm$ refers to the retarded and
advanced sectors of the matrix $E$.

We have thus arrived at a representation of the generating function in terms of
a `flavor' matrix integral involving supermatrices $Z$, $\tilde{Z}$. The Haar
average has erased all microscopic details. Remaining structures are singlet in
Hilbert space,  explaining the overall factor $D$ in the exponent of
Eq.~\eqref{eq:ZZ_sigma_model}. This large factor suggests a semiclassical
evaluation of the matrix integral, which turns out to be exact for the unitary
class.

Before delving into details, let us notice that 
Eq.~\eqref{eq:ZZ_sigma_model} can be brought to the standard form of a non-linear sigma model interpreting $Z$, $\tilde{Z}$ as the rational coordinates of the matrix \cite{Review_of_Sigma_Models}
\begin{align}\label{eq:rational_parametrization}
 Q(Z, \tilde{Z}) &= \begin{pmatrix} 1 & Z \\ \tilde{Z} & 1 \end{pmatrix}
\begin{pmatrix}
1 & 0 \\
0 & -1
\end{pmatrix}
\begin{pmatrix}
1 & Z \\
\tilde{Z} & 1
\end{pmatrix}^{-1} =  T \Lambda T^{-1},
\end{align} 
 fulfilling the non-linear constraint $Q^2=\mathbf{1}_8$.
Expressed in the variable $Q$,  the matrix integral allows for the compact representation
    \begin{align}
\label{eq:generating_function_Q}
     \mathcal{Z} 
     &= 
     \int d(Z, \tilde{Z})\, e^{- D\mathrm{Str} \ln\left( 1 + \frac{1-E}{1+E} Q\Lambda\right)}.
   \end{align}Anticipating that the correlation function of interest only depends on phase differences and that the Fourier transform fixes phases to values $\phi_1 \simeq \phi_2 \simeq \phi_3 \simeq \phi_4\sim {\cal O}(1/D)$, we can linearize Eq.~\eqref{eq:generating_function_Q} in sources and phases, to obtain
\begin{align}
\label{eq:sigma_model_stand}
     \mathcal{Z} 
     &= 
\int dQ\, e^{\frac{iD}{2}{\rm Str}\left(
 \left(\alpha + \phi \right)Q 
\right)}.
\end{align} 
We note that the linearization $e^{i\phi}-1\approx
i\phi$  is a matter of convenience, but not
essential. In time space, this amounts to an approximation of a discrete time
translation operator by a derivative. In line with this observation, the somewhat more elaborate stationary
phase analysis of the un-approximated integral
Eq.~\eqref{eq:generating_function_Q} yields identical results for the SFF,
however then as a function of discrete time. In other words,
Eqs.~\eqref{eq:SFF2dis} and \eqref{eq:SFF2Conn} below are exact, provided $\tau
= t/D $ is considered for $t\in \Bbb{N}$.  This finding is also confirmed by our
numerical experiments.

Proceeding with the linearized formalism, we note that Eq.~\eqref{eq:sigma_model_stand}  is the non-linear sigma model for random
matrix correlations \cite{efetov_book, haake_quantum_signatures_of_chaos}. In
this representation, the matrix degree of freedom Eq.
(\ref{eq:rational_parametrization}) is parametrized as a rotation around the
matrix $\Lambda$, defined in Eq.~\eqref{eq:Lambda}, 
which is known as the `standard saddle' in the present context. A Gaussian expansion in
generators of the rotations $T$ reproduces the diagrammatic perturbation theory
of the previous subsection.  Importantly, the model has additional
`non-perturbative' saddle points. As we discuss next, the latter describes
correlations at longer time scales set by the Heisenberg time.\\

\subsubsection{Semiclassical correlation function}

As mentioned above, the matrix-integral generating (moments of) the SFF in symmetry class A enjoys
`semiclassical exactness'~\cite{haake_quantum_signatures_of_chaos,
Review_of_Sigma_Models, fyodorov, zirnbauer_CUE}: Its evaluation in terms of `saddle points plus
Gaussian fluctuations', provides the exact result.
We thus need to understand the structure of saddle points in addition to 
the standard saddle $\Lambda$.

Referring for further details to Appendix~\ref{app_field_theory}, we notice that causality fixes the bosonic sector of the saddles to take the form
$Q_{\rm bb}=\sigma_3^{ra}\otimes \mathbf{1}_2$. In the fermionic sector, on the other hand, no such restriction applies and any of the sixteen matrices $Q_{\rm ff}={\rm diag}(\pm 1, \pm 1, \pm 1, \pm 1)$ provides a legitimate saddle of the sigma-model. In case of the CUE, however, only 
  the traceless matrices, sharing the signature of the standard saddle point, can be accessed~\footnote{In the sigma model for the CUE, the rational parametrization in Eq. (\ref{eq:rational_parametrization}) fixes the signature of the saddle, since $\tr Q = 0$.  In case of the GUE, however, the constraint $Q^2 = \mathbb{1}_8$ is imposed at the semiclassical level, and one needs to consider saddles with different signatures. In this case,  fluctuations around saddles with finite trace are suppressed by additional factors $1/D$, and thus subleading.}. 
 There are thus $\binom{4}{2} = 6$ relevant saddle points for our calculation. They 
can be parametrized as permutations of the standard saddle $\Lambda$, and classified in terms of the number of involved transpositions $T$, see Table~\ref{tab:saddles_sigma_model}.
\\

\begin{table}[h!]
\centering
\renewcommand{\arraystretch}{1.3} 
\setlength{\tabcolsep}{12pt} 
\begin{tabular}{cccccc}
\hline\hline
$\sigma$ & 1 & 2 & 3 & 4 \\
\hline
$I$ & 1 & 2 & 3 & 4 \\
$T_{13}$ & 3 & 2 & 1 & 4 \\
$T_{14}$ & 4 & 2 & 3 & 1 \\
$T_{23}$ & 1 & 3 & 2 & 4 \\
$T_{24}$ & 1 & 4 & 3 & 2 \\
$T_{14}T_{23}$ & 4 & 3 & 2 & 1 \\
\hline\hline
\end{tabular}
\caption{Six permutations ($\sigma$) giving rise to standard (first entry) and non-standard saddle points (remaining five entries). The last entry corresponds to the Andreev-Altshuler saddle.}
\label{tab:saddles_sigma_model}
\end{table}

A straightforward, if tedious, semiclassical evaluation using the Harish-Chandra-Itzykson-Zuber (HCIZ) formula \cite{correlators_spectra_determinants_quantum_chaos,itzykson_planar_approximation_II, fyodorov}, then leads to 
\begin{widetext}
\begin{equation}\label{eq:semiclassical_Z}
\mathcal{Z}  = \sum_{\sigma} \prod_{i=1,2} \prod_{k=3,4} 
\frac{
(\phi_i - \phi_{\sigma(k)} - \alpha_{\sigma(k)}) (\phi_k - \phi_{\sigma(i)} - \alpha_{\sigma(i)})
}{
(\phi_i - \phi_k) (\phi_{\sigma(i)} - \phi_{\sigma(k)} + \alpha_{\sigma(i)} - \alpha_{\sigma(k)})
}
e^{\frac{iD}{2} \left( \sum_{i=1,2} - \sum_{i=3,4} \right)(\phi_i - \phi_{\sigma(i)} - \alpha_{\sigma(i)})},
\end{equation} 
\end{widetext}
and similar results are obtained for the GUE. 

Next, we compute the  contribution of each saddle to the second moment of the
spectral form factor. Introducing the  compactified notation
$\Delta_{ij}=\phi_i-\phi_j$ and performing the four fold derivative, we get
\begin{widetext}
\begin{equation}\label{eq:G_phi_second_moment}
\begin{aligned}
{\cal F}_I(\phi) &=  \left(\frac{D}{2}\right)^4 - \left(\frac{D}{2}\right)^2 \left(\frac{1}{\Delta_{13}^2} + \frac{1}{\Delta_{23}^2} + \frac{1}{\Delta_{14}^2} + \frac{1}{\Delta_{24}^2}\right)  + \frac{1}{\Delta_{23}^2 \Delta_{14}^2}
+ \frac{1}{\Delta_{13}^2 \Delta_{24}^2},\\
{\cal F}_{T_{13}}(\phi) &=  \left(\frac{D^2}{4\Delta_{24}^2} + \frac{i D}{2\Delta_{13}^2}\left( \frac{1}{\Delta_{12}}+ \frac{1}{\Delta_{23}} + \frac{1}{\Delta_{43} }+ \frac{1}{\Delta_{14}} \right)  - \frac{1}{\Delta_{13}^2 \Delta_{24}^2}  - \frac{1}{\Delta_{12} \Delta_{23} \Delta_{34} \Delta_{41}} \right) e^{iD \Delta_{13}},\\
{\cal F}_{TT}(\phi) &=  \frac{\Delta_{12}^2 \Delta_{34}^2 }{\Delta_{13}^2 \Delta_{32}^2 \Delta_{24}^2 \Delta_{41}^2} e^{iD(\Delta_{13} + \Delta_{24})},
\end{aligned}
\end{equation}
\end{widetext}
the second single transposition following from symmetry. 

Finally, we Fourier transform. Since $\mathcal{Z}$ only depends on phase differences, one of the $\phi_i$ will be trivially integrated out. To fix the redundancy related to the poles with identical causal structure ($\phi_1^+ - \phi_2^+$ and $\phi_3^--\phi_4^-$), we introduce $\phi_s^\pm = \phi_s \pm i \delta_s$ and regulate by enforcing $0<\delta_1 < \delta_2$ and $0<\delta_3 < \delta_4$ (notice that the result is independent of the chosen order). Extending the remaining integrals over $\phi_i$ to infinity, and neglecting contributions involving $\delta$-functions in time,  
we arrive at a disconnected contribution
\begin{align}
    \label{eq:SFF2dis}
\overline{\text{SFF}_{\rm dis}^2}
&=
2D^2~
\begin{dcases}
 \tau^2 \qquad &\text{for }\tau < 1
\\
1 \qquad &\text{for }\tau\geq 1
\end{dcases}\ ,
\end{align}
and connected contribution
\begin{align}
    \label{eq:SFF2Conn}
D\, {\rm Conn}_2(t)
&=
-D ~
\begin{dcases}
~0 \qquad &\text{for }\tau<1/2\\
~2\tau-1 \qquad &\text{for }1<\tau\leq 1/2\\
~1 \qquad &\text{for }\tau\geq 1
\end{dcases}\ ,
\end{align} 
where $\tau = t/D$ is the total time of evolution in units of the Heisenberg time
$\tH=D$ in the CUE. We thus recover the sine-kernel result
Eq.~\eqref{eq_conn2_RMT} of the previous section.

Finally, we comment on the possibility of computing higher moments through the
supersymmetric approach. For the $k$-th moment of the SFF, we need to consider a
$k$-fold replicated theory, which contains a total of $\binom{2k}{k}$ relevant
saddles. The same semiclassical evaluation can be carried out, yielding a
straightforward generalization of Eq.~\eqref{eq:semiclassical_Z}. After taking
$2k$ $\alpha$-derivatives, one obtains $\mathcal{F}_{k}(\phi)$, which can then be Fourier
transformed to yield $\SFF^k$.

To assess the extent of the difficulty of the calculation for higher orders, we look at the example of $k=3$. In this case, the generating function becomes a twelve-dimensional integral over bosonic and fermionic fields, and the saddle point structure includes $\binom{6}{3}=20$ relevant configurations corresponding to traceless sign permutations of the eigenvalue phases. These saddle points generalize those shown in Table \ref{tab:saddles_sigma_model} for $k=2$. As with $k=2$, each non-standard saddle contributes to the connected part ${\rm Conn}_3(t)$, and these contributions scale as one order down in $1/D$ compared to the leading Gaussian contribution $\sim 6 D^3 z(t)^3$.  While the number of relevant saddle point contributions increases combinatorially, their structure is systematically organized by the replica formalism, and the semiclassical exactness of the color-flavor transformation ensures that all such terms are captured. This provides a clear and controlled route to studying the full distribution of the SFF beyond the variance.

\begin{figure}[t]
	\centering
\includegraphics[width=1 \linewidth]{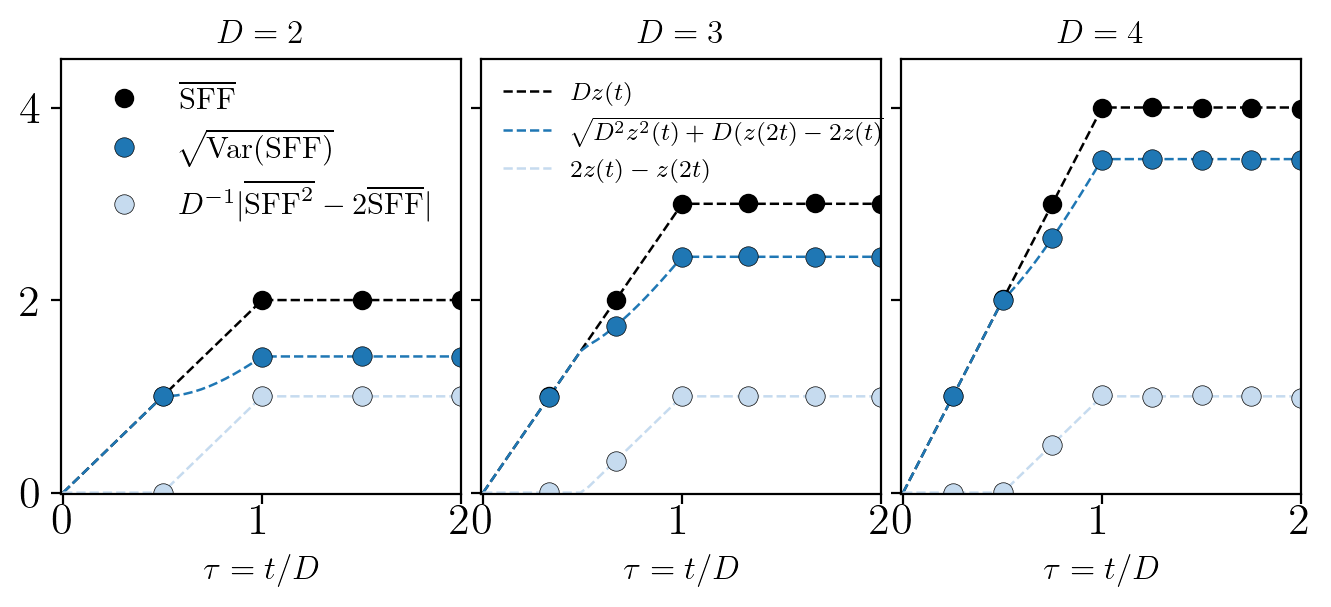}
\caption{Numerical vs. analytical statistics of the SFF statistics of the CUE for very small $D$. Numerical averages (large dots) of $\overline{\rm SFF}$ (black), the variance ${{\rm Var(SFF)}=(\overline{\rm SFF^2}-\overline{\rm SFF}^2)^{1/2}}$ and the deviation from Gaussianity ${D|\overline{\SFF^2}-2\overline {\SFF}^2|}$ are plotted against with the analytical prediction in Eq.\eqref{eq_main}, Eq.\eqref{eq_conn}, (dashed lines) for $D=2,3,4$ in the three panels, respectively. Numerical simulations obtained with $N_{\rm sim}=2\cdot  10^5$.}
	\label{fig_5}
 \vspace{-.5cm}
\end{figure}

Summarizing,  we have obtained the non-Gaussian statistics as a term of next-leading
order in $D^{-1}$ relative to the leading order term, cf. the relative factor of
$D$ between Eqs.~\eqref{eq:SFF2dis} and \eqref{eq:SFF2Conn}. As pointed out
above, these results are exact for the CUE, and as such, in agreement with the
results of  Ref.~\cite{conrey2007howepairssupersymmetryratios}. They are also consistent with
the findings of Ref.~\cite{haakeSommers1999}, which obtained the SFF statistics
in a discrete-time dependent setting via the analysis of  Toeplitz determinants. In this context, the authors of Ref. \cite{haakeSommers1999} further develop the Toeplitz determinant approach by representing the spectral form factor $\abs{t_n^2}=\abs{\tr U^n}^2$
  through determinants of Toeplitz matrices. By expanding the determinant of these matrices
and their logarithm, and evaluating derivatives, they derive \emph{exact} expressions for the mean and variance of SFF. Remarkably, for discrete times $n\leq N/2$ , the moments conform to Gaussian statistics. The plots shown in Fig~\ref{fig_5} provide numerical support for these statements. They show data for the average SFF
and its second moment for ensembles of matrices of very small dimension,
$D=2,3,4$, in perfect agreement with the analytical results. (The dots represent
discrete time steps, the Heisenberg time assuming the values
$t_\mathrm{H}=2,3,4$, respectively.)

It is interesting to note that \emph{all} contributions to
$\textrm{Conn}_2$ are non-perturbative in nature. Even though they are smaller
by just one power $D$ compared to Eq.~\eqref{eq:SFF2dis}, they cannot be obtained from a perturbative, diagrammatic, or semiclassical expansion. For example, in the perturbative
approach outlined in Sec.~\ref{PerturbationTheory}, connected contributions
would assume the form of four `wheels', fully interconnected by ladder `spokes'.
Such diagrams exist, but they are of order $D^{-3}$ relative to the leading
order contributions, and hence negligibly small. The non-perturbative nature of
the supersymmetry results is witnessed by the non-analytic kinks in the time
dependence of $\textrm{Conn}_2$ at $\tau=\frac{1}{2},~1$. They have a status similar to
the celebrated ramp-plateau kink in the average SFF which, likewise, evades
perturbative approaches. 

Finally, the situation with the
GUE is different. There, the background of a non-constant average spectral
density gives rise to a plethora of correction terms of subleading order in
$1/D$. However, these do not affect the finding that $\mathrm{Conn}_2$
Eq.~\eqref{eq_conn2} remains the leading non-Gaussian contribution to the
variance of the SFF.

\section{Conclusion}

In this paper, we studied the statistical moments of the spectral
form factor of circular and Gaussian matrix ensembles. To leading order in the
matrix dimension, $D$, the SFF is Gaussian with deviations appearing at next leading order.
Employing two complementary approaches, sine-kernel techniques and
supersymmetric matrix integration, we identified a universal subleading
correction to the second moment, consistent with previous findings for the
circular ensemble~\cite{haakeSommers1999}. Our sine-kernel analysis clarified discrepancies with previous studies, while the
field-theoretic analysis revealed new non-perturbative saddle points that extend
beyond the standard ones in the non-linear sigma model. These findings were
further corroborated by numerical simulations, which confirmed the theoretical
predictions.

Our results open several avenues for further exploration. Alongside the
generalization to other symmetry classes, one is the
extension of our approach to the computation of higher-order subleading
corrections to the SFF, or even its probability distribution.  Another interesting question is to what degree the subleading corrections to the SFF are 
universal across different chaotic systems beyond Random Matrix Theory. 
Our focus has been on RMT ensembles as a controlled setting to explore subleading corrections to the spectral form factor; however, it is natural to ask whether such effects are observable in concrete quantum many-body systems. While these corrections may be subtle and challenging to detect directly in physical observables, their universality makes them valuable as benchmarks for diagnosing quantum chaoticity in many-body systems. In particular, moments of the spectral form factor naturally arise in the analysis of time-dependent observables such as correlation functions (including out-of-time-order ones) and return probabilities. As such, the random matrix theory predictions provide a baseline upon which one can analyze the influence of the specific many-body spectrum in realistic systems. 
One promising direction is provided by the Sachdev-Ye-Kitaev (SYK) model, where finite-$N$ corrections lead to subleading deviations from random matrix predictions, including in the statistics of the SFF.

\begin{acknowledgments} 
\textit{Acknowledgments.} SP thanks A. De Luca for helpful exchanges regarding
approximated Gaussian behavior, AA thanks L. Erdoš for enlightening
discussions concerning the rigor of different approaches to spectral statistics and FD thanks D. Rosa for helpful discussions and related projects. We acknowledge funding by the Deutsche Forschungsgemeinschaft (DFG, German Research
Foundation) under Projektnummer 277101999 - TRR 183 (project B02 and
A03), and under Germany's Excellence Strategy - Cluster of Excellence
Matter and Light for Quantum Computing (ML4Q) EXC 2004/1 - 390534769 and financial support by Brazilian agencies CNPq and FAPERJ. FD would like to thank the Perimeter/ICTP-SAIFR/IFT-UNESP fellowship program and CAPES for financial support. Research at Perimeter Institute is supported in part by the Government of Canada through the Department of Innovation, Science and Economic Development and by the Province of Ontario through the Ministry of Colleges and Universities.

\emph{Data availability.}
The data underlying the figures and the source code used for generating the simulations are available at \cite{zenodo}.

\end{acknowledgments}

\newpage
\appendix
\onecolumngrid
\section{Sine-Kernel predictions for Gaussian unitary ensemble}
\label{app_sinekernel}

In this appendix, we seek to implement the sine-kernel formalism outlined in Sec.~\ref{sub_sign-kernel} to compute the second moment of the spectral form factor of the GUE ensemble. We adhere closely to the framework established in \cite{Liu2019spectral} while rectifying critical inconsistencies, which lead to a different result for the second moment of the SFF. \\

Equation Eq.~\eqref{eq:rho_n} demonstrates how all higher-order correlations can be expressed as powers of the sine-kernel. To transition from this expression to the spectral form factor in Eq. \eqref{eq_sffn}, it is necessary to apply the appropriate Fourier transform. For this purpose, the following \emph{convolution formula} [cf. Eq. (6.2.18) in \cite{mehta2004random}], valid in the large $D$ limit,  proves particularly useful:
\begin{align}
\begin{split}
     \int \prod_{i=1}^{m} d\lambda_i &\, K(\lambda_1, \lambda_2) K(\lambda_2, \lambda_3) \cdots K(\lambda_m, \lambda_1) e^{i \sum_{i=1}^{m} k_i \lambda_i} \\ 
   &=  \delta\left(\sum_{i=1}^{m} k_i\right) \int dk \, g(k)\prod_{i=1}^{m-1} g(k + \sum_{l=1}^i\frac{k_l}{2D})\ ,
    \label{eq:convolution_formula}
    \end{split}
\end{align}
where the function $g(k)$ is defined as:
\begin{equation}
   g(k) = 
    \begin{cases}
        1 & \text{for } |k| < \frac{1}{2}, \\
        0 & \text{for } |k| > \frac{1}{2} \ .
    \end{cases}
    \label{eq:fourier_transform_sine_kernel}
\end{equation}
We note that the omission of the sum over $k_l$   in Eq. (2.36) of \cite{Liu2019spectral} is partially responsible for the errors that propagate through the subsequent computations. \\
Let us here clarify the differences with the notations of Ref.~\cite{Liu2019spectral}, which define the moments of the SFF in terms of some functions $r_k(t)$ with $k=1,2,3$.
These are related to our notations by $r_2(t) = 1-z(t)$ and $D r_1(t) = \bar u_t = \int_{\lambda} K(\lambda_i,\lambda_i) e^{i \lambda_i t} $. Finally, we neglect  $r_3(t)$, corresponding to the finite size correction of the delta function in Eq. \eqref{eq:convolution_formula} above.\\

To evaluate $\overline{\SFF^2}$ for GUE, here we proceed by computing the first and second terms of Eq. (\ref{eq:R4_total}) in the following.

\subsubsection{Third order term}
 The term $ \int_\lambda  R_3(\lambda_1,\lambda_2, \lambda_3) e^{i(2\lambda_1 - \lambda_2-\lambda_3)t}$,  expressed in terms of the determinant of the sine kernel using Eq. \eqref{eq:rho_n} results in a multiple number of terms that need to be computed through Eq.~(\ref{eq:kernel_K}) and Eq.~(\ref{eq:convolution_formula}). Due to the large number and tedious nature of the computation of these terms, stating the calculation of each term is beyond the scope of this discussion. For those, we refer the reader to Ref.~\cite{Liu2019spectral}, while here we only explicitly adjust the inconsistencies in Eq.~ (2.68) (the so-called 3-type contribution), namely:

\begin{equation}
 \begin{split}
      2 \Re  \int_\lambda \, K(\lambda_1, \lambda_2) K(\lambda_2, \lambda_3) K(\lambda_3, \lambda_1) e^{i(2\lambda_1 - \lambda_2 - \lambda_3)t}
      =2D (1-z(2t)) \ .
\label{eq:3_type_integral}
        \end{split}
    \end{equation}

Considering this correction we arrive at Eq. \eqref{eq: R_3}. \\

\subsubsection{Fourth-order term}
A similar approach is taken for the integral
 $\int_\lambda R_4(\lambda_1,\lambda_2, \lambda_3, \lambda_4)e^{i(\lambda_1 + \lambda_2-\lambda_3-\lambda_4)t}$, which is rewritten in terms of the sine kernel determinant. Following Ref.~\cite{Liu2019spectral} and correcting Eq.~(2.56) (the so-called 4-type contribution), we derive:
  
\begin{align}
\begin{split}
    \int_\lambda  K(\lambda_1, \lambda_3) K(\lambda_3, \lambda_2) K(\lambda_2, \lambda_4) K(\lambda_4, \lambda_1) e^{i(\lambda_1 + \lambda_2 - \lambda_3 - \lambda_4)t}=D (1-z(t)) \ ,
    \end{split}
    \end{align}
\begin{align}
\begin{split}
    \int_\lambda K(\lambda_1, \lambda_2) K(\lambda_2, \lambda_3) K(\lambda_3, \lambda_4) K(\lambda_4, \lambda_1) e^{i(\lambda_1 + \lambda_2 - \lambda_3 - \lambda_4)t}=D (1-z(2t)) \ ,
    \end{split}
    \end{align}
 \begin{align}
 \begin{split}
         \int_\lambda K(\lambda_1, \lambda_2) K(\lambda_2, \lambda_4) K(\lambda_4, \lambda_3) K(\lambda_3, \lambda_1) e^{i(\lambda_1 + \lambda_2 - \lambda_3 - \lambda_4)t}=D (1-z(2t)) \ .
           \end{split}
\end{align}
By incorporating this correction, we obtain Eq. \eqref{eq: R_4}.

\subsubsection{The final result}

By aggregating the previously derived results along with the two-point function contributions from the third and fourth terms of Eq. \eqref{eq:R4_total}, we arrive at Eq. \eqref{eq:R4_final_result}. To obtain the corresponding expression for the GUE, we substitute  $z(t)$ with Eq. \eqref{eq_brezin}.

\section{Field-theory details}\label{app_field_theory}
\subsection{Matrix action}
In this appendix, we provide additional details on the derivation of the effective field theory discussed in Sec.~\ref{sec_fieldtheory}. Our starting point is the generating function
\begin{equation}
    \mathcal{Z}(\alpha, \phi) = \Bigg\langle \frac{\det( 1 - e^{i(\alpha_1 + \phi_1^+)}U^\dagger) \det( 1 - e^{i(\alpha_2 + \phi_2^+)} U^\dagger)\det( 1 - e^{-i(\alpha_3 + \phi_3^-)}U) \det( 1 - e^{-i(\alpha_4 + \phi_4^-)} U)}{\det( 1 - e^{+i \phi_1^+}U^\dagger) \det( 1 - e^{+ i \phi_2^+} U^\dagger)\det( 1 - e^{-i \phi_3^-}U) \det( 1 - e^{-i \phi_4^-} U)} \Bigg\rangle_U,
\end{equation}
involving Haar average of the ratio of eight spectral determinants. The latter  
can be expressed in terms of integrals over $4D$ fermionic variables $\chi_a^i$ and $4D$ bosonic variables $S_a^i$, where $i = 1, \dots, D$ and $a = 1,2,3,4$.  To compactify notation, it is convenient to introduce the $8D$-component  supervector $\psi$ 
\begin{equation}
    \psi_a^i = \begin{pmatrix}
        \chi_a^i\\ S_a^i
    \end{pmatrix},
\end{equation}
 comprising of eight internal degrees of freedom, viz the two `graded' fermionic/bosonic components (${\rm b/f}$), the two `causal' retarded/advanced components ($+/-$), and the two replica components ($1/2$).
The ensemble average of the generating function can then be expressed as
\begin{equation}
\mathcal{Z} =   \int_{\mathrm{U}(D)}\mkern-25mu dU \int \! d(\psi, \bar\psi) ~ \exp \biggr( \bar \psi_{+}  (1-E_+ U^\dagger) \psi_{+}+ \bar \psi_{-}  (1-E_- U) \psi_{-}\biggr).
\end{equation}where $E_\pm$ denote the retarded/advanced sectors of $E$, defined in Eq. (\ref{eq:def_E}).

The next step is to perform the integral over the Haar ensemble employing the color-flavor transformation  identity \cite{haake_quantum_signatures_of_chaos, Review_of_Sigma_Models}
\begin{equation}
    \int_{\mathrm{U}(D)}\mkern-25mu dU ~ \exp \biggr( \bar \psi_{1}  U \psi_{2'} + \bar \psi_{2} U^\dagger \psi_{1'}\biggr) = \int d(\tilde{Z},Z) ~ \text{Sdet}(1-\tilde{Z}Z)^D \exp \biggr( \bar \psi_{1}  \tilde{Z} \psi_{1'} + \bar \psi_{2} Z \psi_{2'}\biggr),
\end{equation}where `$\text{Sdet}$' denotes the superdeterminant \cite{haake_quantum_signatures_of_chaos}. We are then left with a quadratic action in the superfield $\psi$ that can be  integrated out, resulting in
\begin{equation*}
\begin{aligned}
    \mathcal{Z} &= \int d(Z, \tilde{Z})~d(\psi, \bar \psi)~  \mathrm{Sdet}(1 - \tilde{Z} Z)^D \exp(\bar \psi M\psi)
                = \int d(Z, \tilde{Z})~\mathrm{Sdet}(1 - \tilde{Z} Z)^D~\mathrm{Sdet}(M)^{-D},
                \end{aligned}
\end{equation*}where $M = \left(\begin{smallmatrix} 1 & E_+\tilde{Z} \\ E_-Z & 1 \end{smallmatrix}\right)$ is the structure in causal indices. Evaluating the determinant in causal indices and using that $\ln ~\text{Sdet} = \text{Str} ~\ln$, we arrive at Eq. (\ref{eq:ZZ_sigma_model}) in the main text. To rewrite this in the standard form of a sigma model, we follow \cite{Review_of_Sigma_Models} and introduce
    \begin{align}
    \label{eq:sigma_model_Q}
        Q(Z, \tilde{Z}) =   T \Lambda T^{-1},\qquad T =  \begin{pmatrix} 1 & Z \\ \tilde{Z} & 1 \end{pmatrix},
        \end{align}
and in terms of which 
\begin{align}
S[Q] = D~\mathrm{Str} \ln( 1 + \frac{1-E}{1+E} Q \Lambda ),
\end{align}
with $\text{Str}A = \tr A_{BB} -\tr A_{FF} $. Notice that $Q^2=\mathbf{1}_8$, so 
Eq.~(\ref{eq:sigma_model_Q}) indeed corresponds to a non-linear sigma model.

\subsection{Semiclassical evaluation}

For the semiclassical evaluation, it is convenient to start out from an alternative matrix action, involving a quadratic contribution $\frac{D}{2}\text{Str} (Q^2)$ which enforces the nonlinear constraint as a result of the saddle approximation. Specifically, the sigma model in the main text is semiclassically equivalent to the matrix integral
\begin{equation}\label{eq:app_semiclassically_equiv_matrix_integral}
   \mathcal{Z}= \int dQ~ \exp\left( \frac{D}{2} \text{Str}(Q^2) - D ~\text{Str} \ln\left(Q + \frac{i}{2} (\alpha + \phi) \right) \right),
\end{equation}
where the integration now is not restricted to the manifold $Q^2=\mathbf{1}_8$. 

Now, we are allowed to shift $Q \to Q - \frac{i}{2}(\alpha + \phi)$, making the logarithm rotationally invariant. This then motivates the introduction of  angle and radial coordinates
\begin{equation}
A = VRV^{-1}, \quad R = \text{diag}(z_1, z_2, z_3, z_4, w_1, w_2, w_3, w_4),
\end{equation} 
where rotations $V$ contain the angles and the integration contour of radial coordinates is over the real axis for the $w$ variables (fermionic sector) and over the imaginary axis for the $z$ variables (bosonic sector). The change of integration variables introduces a Bereznian $B = \frac{\Delta^2(z,w)}{\Delta^2(z)\Delta^2(w)}$ given by the Vandermonde determinants
\begin{equation}
\begin{aligned}
\Delta(x) = \prod_{i<j} (x_i - x_j), \qquad \qquad \Delta(x, y) = \prod_{i,k} (x_i - y_k).
\end{aligned}
\end{equation} 
The integration over angles $V$ is non-trivial but can be carried out employing the Harish-Chandra-Itzykson-Zuber (HCIZ) integration formula \cite{correlators_spectra_determinants_quantum_chaos,itzykson_planar_approximation_II}, leading to
\begin{equation}
\mathcal{Z} = \frac{\Delta(\phi, \phi + \alpha)}{\Delta(\phi) \Delta(\phi + \alpha)} \int dZ ~dW ~ \frac{\Delta(z) \Delta(w)}{\Delta(z, w)} ~ \exp\left(\frac{D}{2} \mathrm{Str} \left(R - \frac{i}{2}(\alpha + \phi)\right)^2 -  D ~\mathrm{Str} \ln R \right).
\end{equation}
Shifting back $R \to R +\frac{i}{2}(\alpha + \phi)$ and performing the saddle point analysis, one finds that the integration contours over the bosonic block variables can only be deformed to reach the saddle points $z_{1,2} = +1$, $z_{3,4} = -1$, while the contours over the fermionic block variables are allowed to reach any of the saddles $w_i = \pm 1$. While in the sigma model Eq.~\eqref{eq:rational_parametrization} $\tr Q = 0$ and the signature of $Q$ is fixed, this is not the case in Eq.~\eqref{eq:app_semiclassically_equiv_matrix_integral} and all 16 saddles must be considered. However, fluctuations around saddles with finite trace are suppressed by additional factors $1/D$, and are thus subleading~\cite{haake_quantum_signatures_of_chaos}. In any case, only $6$ saddles are relevant for our calculation, so we arrive at 
\begin{equation}\label{eq:Z_before_simplifying_Delta}
\mathcal{Z} = \frac{\Delta(\phi, \phi + \alpha)}{\Delta(\phi) \Delta(\phi + \alpha)} \sum_P \frac{\Delta(z - i\phi/2) \Delta(w - i\phi/2 - i\alpha/2) }{\Delta(z - i\phi/2, w - i\phi/2 - i\alpha/2)} e^{i\frac{D}{2} \mathrm{Str} R(\alpha + \phi)},
\end{equation}
with $\sum_P$, the sum over the 6 saddle points. As discussed in the main text, they can be expressed as permutations of the standard saddle $ P_0 \equiv \Lambda $ and classified according to the number of transpositions $T$, see Table~\ref{tab:saddles_sigma_model}, resulting in  
\begin{equation}
\mathcal{Z} = \sum_{\sigma} \prod_{i=1,2} \prod_{k=3,4} 
\frac{
(\phi_i - \phi_{\sigma(k)} - \alpha_{\sigma(k)}) (\phi_k - \phi_{\sigma(i)} - \alpha_{\sigma(i)})
}{
(\phi_i - \phi_k) (\phi_{\sigma(i)} - \phi_{\sigma(k)} + \alpha_{\sigma(i)} - \alpha_{\sigma(k)})
}
e^{i \frac{D}{2} \left( \sum_{i=1,2} - \sum_{i=3,4} \right)(\phi_i - \phi_{\sigma(i)} - \alpha_{\sigma(i)})}.
\end{equation}

\newpage
\bibliography{biblio}

\end{document}